\documentclass[11pt,a4paper]{article}
\usepackage{jheppub}
\usepackage{feynmp}
\usepackage{epstopdf}
\usepackage{caption}          % for subfigures
\usepackage{subcaption}       % for subfigures
\def\be{\begin{equation}}
\def\ee{\end{equation}}
\def\bea{\begin{eqnarray}}
\def\eea{\end{eqnarray}}
\def\beal{\begin{equation}\begin{aligned}}
\def\eeal{\end{aligned}\end{equation}}
\def\nn{\nonumber}

\def\bra#1{\langle #1|}

\def\braket#1{\langle #1 \rangle}
\def\u#1{\underline{#1}}
\def\o#1{\overline{#1}}

\def\Res_#1{\operatorname*{Res}_{#1}}
\def\Tr{\operatorname*{Tr}}

\def\tf{\tilde{f}}
\def\shuffle{\sqcup\mathchoice{\mkern-7mu}{\mkern-7mu}{\mkern-3.2mu}{\mkern-3.8mu}\sqcup}

\def\ie{i.e. }
\def\eg{e.g. }

\def\eqn#1{eq.~\eqref{#1}}
\def\eqns#1#2{eqs.~\eqref{#1} and~\eqref{#2}}
\def\Eqn#1{Eq.~\eqref{#1}}
\def\Eqns#1#2{Eqs.~\eqref{#1} and~\eqref{#2}}
\def\fig#1{figure~{\ref{#1}}}

\def\tab#1{table~{\ref{#1}}}

\def\sec#1{section~{\ref{#1}}}

\def\app#1{appendix~{\ref{#1}}}

\def\rcite#1{ref.~\cite{#1}}
\def\rcites#1{refs.~\cite{#1}}

\def\citeColorOrdering{\cite{Berends:1987cv,Mangano:1987xk,Mangano:1988kk,Bern:1990ux}}
\def\citeUnitarityMethod{\cite{Bern:1994zx,Bern:1994cg,Bern:1996je}}

\def\fmfsettings{
    \fmfset{thin}{1.2pt}
    \fmfset{arrow_len}{7pt}
    \fmfset{arrow_ang}{20}
    \fmfset{wiggly_len}{7pt}
    \fmfset{curly_len}{6pt}
    \fmfset{dash_len}{8pt}
    \fmfset{dot_len}{3pt}
}

\title{Color-Kinematics Duality for QCD Amplitudes}
\author[a,b,c]{Henrik Johansson,}
\author[d]{Alexander Ochirov}

\affiliation[a]{Theory Division, Physics Department, CERN, CH--1211 Geneva 23, Switzerland}
\affiliation[b]{Department of Physics and Astronomy, Uppsala University, Box 516, SE-75120 Uppsala, Sweden}
\affiliation[c]{Nordita, KTH Royal Institute of Technology and Stockholm University, Roslagstullsbacken 23, SE-10691 Stockholm, Sweden}
\affiliation[d]{Higgs Centre for Theoretical Physics, School of Physics and Astronomy,\\ The University of Edinburgh, Edinburgh EH9 3JZ, Scotland, UK}

\emailAdd{henrik.johansson@physics.uu.se}
\emailAdd{alexander.ochirov@ed.ac.uk}

\abstract{We show that color-kinematics duality is present in tree-level amplitudes of quantum chromodynamics with massive flavored quarks.
Starting with the color structure of QCD, we work out a new color decomposition for $n$-point tree amplitudes in a reduced basis of primitive amplitudes.
These primitives, with $k$ quark-antiquark pairs and $(n-2k)$ gluons, are taken in the $(n-2)!/k!$ Melia basis, and are independent under the color-algebra Kleiss-Kuijf relations. This generalizes the color decomposition of Del Duca, Dixon, and Maltoni to an arbitrary number of quarks.
The color coefficients in the new decomposition are given by compact expressions valid for arbitrary gauge group and representation.
Considering the kinematic structure, we show through explicit calculations that color-kinematics duality holds for amplitudes with general configurations of gluons and massive quarks.
The new (massive) amplitude relations
that follow from the duality can be mapped to a well-defined subset of the familiar BCJ relations for gluons.
They restrict the amplitude basis further down to $(n-3)!(2k-2)/k!$ primitives,
for two or more quark lines.
We give a decomposition of the full amplitude in that basis.
The presented results provide strong evidence that QCD obeys the color-kinematics duality, at least at tree level.
The results are also applicable to supersymmetric and $D$-dimensional extensions of QCD.}

\preprint{CERN-PH-TH-2015-149 \\
\phantom{~} \hfill UUITP-13/15 \\
\phantom{~} \hfill NORDITA-2015-79 \\
\phantom{~} \hfill Edinburgh 2015/11}

\begin{document}
\maketitle

\pagebreak
%%%%%%%%%%%%%%%%%%%%%%%%%%%%%%%%%%%%%%%%%%%%%%%%%%%%
\section{Introduction}
\label{sec:intro}
%%%%%%%%%%%%%%%%%%%%%%%%%%%%%%%%%%%%%%%%%%%%%%%%%%%%

The sheer abundance of QCD background processes at the Large Hadron Collider has stimulated remarkable theoretical progress in the treatment of perturbative gauge theory. Calculations of previously intractable processes involving high numbers of loops and legs have been successfully carried out,
such as the inclusive $W+5$-jet production at next-to-leading order~\cite{Bern:2013gka}, or the inclusive Higgs production
at (next-to)$^3$-leading order~\cite{Anastasiou:2015ema}. Modern methods have been essential for many aspects of this progress, such as the unitarity method~\citeUnitarityMethod, on-shell recursion~\cite{Britto:2004ap,Britto:2005fq}, more recent unitarity-based one-loop methods~\cite{Ellis:2007br,Berger:2008sj,Ossola:2007ax,Mastrolia:2008jb,
Giele:2008bc,Ellis:2008qc,Berger:2009zg,Bevilacqua:2009zn,
Mastrolia:2010nb,Berger:2010zx,Badger:2010nx,Hirschi:2011pa}, and advanced integration techniques~\cite{Gehrmann:1999as,Anastasiou:2002yz,Anastasiou:2003yy,Smirnov:2004ym,Duhr:2011zq,Anastasiou:2013srw,Caola:2014lpa}, to mention a few.
Even so, increasingly refined methods are needed to curb the factorial growth of complexity characteristic of perturbative computations. 

The increase of both the number of loops and legs is an unavoidable consequence of precision QCD. In the physical observables, contributions from different numbers of loops and legs are tied together through the singularity structure and unitarity. 
High-multiplicity tree amplitudes enter explicitly in the real emission contributions, as well as secretly in the virtual loop contributions through the modern unitarity-based methods. Therefore, progress in treatment of tree amplitudes can be directly or indirectly translated to progress in precision QCD phenomenology.

Modern methods usually decompose tree amplitudes into purely-kinematic primitive amplitudes and their coefficients that depend only on color.
Such primitive amplitudes are gauge-invariant and are often computed via off-shell~\cite{Berends:1987me} or on-shell recursion~\cite{Britto:2004ap,Britto:2005fq}, which admits an analytic all-multiplicity solution~\cite{Drummond:2008cr,Dixon:2010ik} for massless QCD with up to four quark-antiquark pairs. The primitives can be assembled in several superficially different ways, with varying degree of efficiency, to obtain the full amplitude. The standard SU$(N_c)$ trace-based decomposition of an $n$-gluon amplitude involves
an overcomplete summation over $(n-1)!/2$ linearly-dependent primitives.
The linear dependence comes from two sources: from the color algebra of the gauge theory~\cite{Kleiss:1988ne} and from an observed kinematic algebra,
which is closely tied to the former through the color-kinematics duality~\cite{Bern:2008qj,Bern:2010ue}. The corresponding relations satisfied by the primitives are known to all multiplicity: the Kleiss-Kuijf (KK) relations~\cite{Kleiss:1988ne}, and the Bern-Carrasco-Johansson (BCJ) relations~\cite{Bern:2008qj}.

More efficient ways of assembling the primitives are known. The Del Duca-Dixon-Maltoni (DDM)~\cite{DelDuca:1999ha,DelDuca:1999rs} color decomposition removes all the pure-gluon primitives that are redundant under the KK relations, and thus involves only the $(n-2)!$ members of the KK primitive basis. For amplitudes with a single quark-antiquark pair a very similar decomposition can be used~\cite{Kosower:1987ic,Mangano:1988kk}. For general QCD amplitudes, with $k$ quark-antiquark pairs and $(n-2k)$ gluons, much less is known about non-redundant decompositions involving well-defined and simple primitive amplitudes. Recently, Melia proposed~\cite{Melia:2013bta,Melia:2013epa} a basis of primitives, for general $n$ and $k$, that enjoys the same useful properties as the KK basis. For distinctly-flavored quarks the basis consists of $(n-2)!/k!$ planar color-ordered amplitudes. In this paper we give a color decomposition that uses precisely these primitives. Thus we obtain a generalization of the DDM decomposition valid for all tree amplitudes in QCD, applicable to any gauge group and any representation.

The color-kinematics duality~\cite{Bern:2008qj,Bern:2010yg} highlights the fact that gauge theories generically possess an underlying kinematic structure that controls the theory, similar to the way the color Lie algebra defines the theory. This structure has been used successfully for tree and loop amplitudes in massless gauge theories with and without supersymmetry, including massless QCD with no quarks~\cite{Bern:2010ue,Carrasco:2011mn,Bern:2011rj,Bern:2012uf,Carrasco:2012ca,Boels:2013bi,Bjerrum-Bohr:2013iza,Bern:2013yya,Nohle:2013bfa,Bern:2013uka,Chiodaroli:2013upa,Johansson:2014zca,Bern:2014sna}. The duality imposes the BCJ relations on pure-gluon tree amplitudes that further reduce the basis of independent primitives to one of size $(n-3)!$. These relations have been observed to still hold in certain specific excursions away from the massless pure-gluon case, \eg for one massless quark-antiquark pair~\cite{Sondergaard:2009za,Weinzierl:2014ava}, or for two or three massive particles~\cite{Naculich:2014naa,Naculich:2015zha}.
The color-kinematics duality was extended to theories with fundamental massless matter in refs.~\cite{Johansson:2014zca,Chiodaroli:2013upa}, but the corresponding BCJ relations were not worked out there.
They are instead presented here for matter with any mass.

In this paper, we show that the color-kinematics duality can be straightforwardly extended to all tree-level amplitudes in QCD. Furthermore, for amplitudes with $k$ massive distinctly-flavored quark-antiquark pairs and $(n-2k)$ gluons, we find all BCJ relations generated by the duality. For $k\ge2$ they reduce the number of independent primitives to $(n-3)!(2k-2)/k!$, and for $k=0,1$ the counting is the standard $(n-3)!$. Using this reduced BCJ basis, we construct a second new decomposition that uses these primitives only.
We expect that the BCJ relations obtained here are applicable beyond the case of QCD. In particular, super-QCD and $D$-dimensional QCD should obey the color-kinematics duality at tree level just as well.

This paper is organized in an example-driven way.
After we review the color and kinematic structure of gauge theory amplitudes
in \sec{sec:review}, we proceed to the lower-point examples
that lead to the main result of \sec{sec:coloralgebra} --
the new color decomposition for a generic QCD amplitude
(\sec{sec:colorfactors}).
We return to the lower-point examples in the beginning of
\sec{sec:kinematicalgebra},
where we show that they respect the color-kinematics duality
and thus satisfy the massive BCJ relations. Then we generalize them
and derive a new amplitude basis in \sec{sec:bcjsolution}.
Our final amplitude decomposition is given in \sec{sec:ckfactors}.
We conclude by discussing our results in \sec{sec:conclusion}.

%%%%%%%%%%%%%%%%%%%%%%%%%%%%%%%%%%%%%%%%%%%%%%%%%%%%
\section{Review and preliminaries}
\label{sec:review}
%%%%%%%%%%%%%%%%%%%%%%%%%%%%%%%%%%%%%%%%%%%%%%%%%%%%

In this section we review some general properties of
the color and kinematic structure of tree-level scattering amplitudes in QCD.\footnote{In this paper by QCD we mean Yang-Mills theory with gauge group $G$ and with $N_f$ massive Dirac fermions in the fundamental representation (quarks). Additionally, we allow for supersymmetric extensions. The cases of ${\cal N}=0,1,2$ (super-)QCD are all included in the general treatment that follows.}
Amplitudes involving only gluons or at most one quark-antiquark pair
have a similar form and structure, and are well studied in the literature.
Adding more quarks makes the composition of the amplitude more involved, and this is the topic of the bulk of this paper.

QCD is a renormalizable gauge theory with only cubic and quartic interactions.
In what follows we center the discussion around the cubic interactions.
The role of the quartic ones is to make the amplitudes
constructed from the Feynman rules gauge invariant,
and they carry no new physical information with respect to the cubic interactions.
This nontrivial statement is made apparent
by the on-shell recursion~\cite{Britto:2004ap,Britto:2005fq},
which relies only on input from the three-point amplitudes of the theory.

For the color structures the redundancy of quartic interactions is clear from inspecting the four-gluon Feynman vertex
\be
\parbox{45pt}{
\begin{fmffile}{g4vertexfull} \fmfframe(0,10)(0,10){
\fmfset{curly_len}{5pt}
\begin{fmfgraph*}(30,33)
      \fmflabel{$a,\lambda\!$}{g1}
      \fmflabel{$b,\mu$}{g2}
      \fmflabel{$c,\nu$}{g3}
      \fmflabel{$d,\rho$}{g4}
      \fmftop{g2,g3}
      \fmfbottom{g1,g4}
      \fmf{curly}{v,g1}
      \fmf{curly}{v,g2}
      \fmf{curly}{v,g3}
      \fmf{curly}{v,g4}
      \end{fmfgraph*} }
      \end{fmffile} }
 \, = \begin{aligned} \hspace{270pt} \\
      \frac{ig^2}{2}
      \big[ \tf^{abe} \tf^{ecd}
            \big( g^{\lambda \nu} g^{\mu \rho}
                - g^{\lambda \rho} g^{\mu \nu} \big)
          + \tf^{bce} \tf^{eda}
            \big( g^{\lambda \nu} g^{\mu \rho}
                - g^{\lambda \mu} g^{\nu \rho} \big) & \\
          + \tf^{ace} \tf^{ebd}
            \big( g^{\lambda \mu} g^{\nu \rho}
                - g^{\lambda \rho} g^{\mu \nu} \big) &
      \big] \,,
      \end{aligned} 
\ee
where we, for later convenience, use imaginary structure constants $\tf^{abc} = i\sqrt{2} f^{abc}$.
This vertex contains the same color factors as the  $s$-, $t$- and $u$-channel diagrams that are constructed from the three-gluon vertices.
Hence the quartic vertex can always be absorbed in some way into cubic trees
without changing the general color structure of a QCD amplitude.

Therefore, without loss of generality we can write a QCD tree amplitude in terms of cubic graphs only. This gives us an expansion of the $n$-point tree amplitude of the form\footnote{Following \rcite{Bern:2008qj} we absorb all factors of $i$ into the numerators, which is convenient for tree amplitudes. The numerators in \rcite{Johansson:2014zca} have a factor of $-i$ pulled out relative to this convention.}
\be
   {\cal A}^{\text{tree}}_{n,k} = g^{n-2}\!\!\!
      \sum_{\text{cubic graphs }\Gamma_i}^{\nu(n,k)} \! \frac{c_i n_i}{D_i} \,,
\label{BCJformYM}
\ee
where $c_i$ are color factors, $n_i$ are kinematic numerators, and
$D_i$ are denominators encoding the propagator structure of the cubic graphs.
The denominators may contain masses,
corresponding to massive quark propagators.

%%%%%%%%%% TABLE %%%%%%%%%%
\begin{table*}[t]
\centering
\begin{tabular}{|c||c|c|c|c|c|c|}
\hline
$k \setminus n$ & 3 & 4 & 5 & 6 & 7 & 8 \\
\noalign{\hrule height 0.7pt}
0 & 1 & 3 & 15 & 105 & 945 & 10395 \\
\hline
1 & 1 & 3 & 15 & 105 & 945 & 10395 \\
\hline
2 & - & 1 &  5 &  35 & 315 &  3465 \\
\hline
3 & - & - &  - &   7 &  63 &   693 \\
\hline
4 & - & - &  - &   - &   - &    99 \\
\hline
\end{tabular}
\\ \vspace{8pt}
$\nu(n,k)=\frac{(2 n-5)!!}{(2 k-1)!!}$ for $2k \le n$
\\ \vspace{-7pt}
\caption{\small Number of cubic graphs, $\nu(n,k)$,
         in the full $n$-point amplitude with $k$ distinguishable quark-antiquark pairs
         and $(n-2k)$ gluons.}
\label{tab:CubicGraphsDistinguishable}
\end{table*}
%%%%%%%%%%%%%%%%%%%%%%%%%%%

Amplitudes with multiple quarks of the same flavor and mass can be obtained
from distinct-flavor amplitudes
by setting masses to be equal and summing over permutations of quarks.
Therefore, we do not lose generality
by taking all $k$ quark-antiquark pairs to have distinct flavor and mass.
For explicitness, in \tab{tab:CubicGraphsDistinguishable}
we provide total counts of cubic graphs for different amplitudes
up to eight particles and four quark pairs.
This count appears in \eqn{BCJformYM} as $\nu(n,k)$. It agrees with the usual counting of standard QCD Feynman diagrams restricted to those diagrams that only have trivalent vertices.

The kinematic numerators $n_i$ are in general not gauge-invariant and thus have no unique expressions. However, certain linear combinations of $n_i/D_i$ correspond to gauge-invariant primitive amplitudes that can be constructed from the color-stripped Feynman rules~\cite{Dixon:1996wi} summarized in \app{app:FeynmanRules}.

\begin{figure}[t]
\centering
\includegraphics[scale=1.0,trim=0 0 0 0,clip=true]{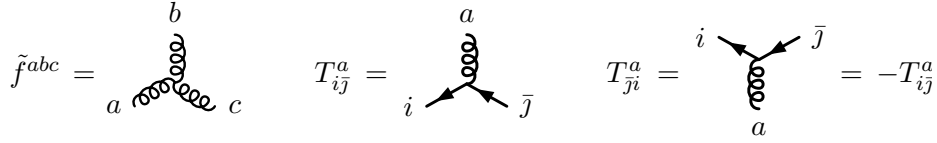}
\vspace{-7pt}
\caption{\small Color vertices with planar ordering
         consistent with the color-stripped Feynman rules.}
\label{fig:colorvertices}
\end{figure}

The color factors $c_i$ in \eqn{BCJformYM} are constructed
from the cubic graphs using only two building blocks:
the structure constants $\tf^{abc}$ for three-gluon vertices
and generators $T^a_{i \bar \jmath}$ for quark-gluon vertices,
as shown in \fig{fig:colorvertices}.
When separating the color from kinematics,
the diagrammatic crossing symmetry only holds up to signs
dependent on the permutation of legs.
These signs are apparent in the total antisymmetry of $\tf^{abc}$.
Since the structure constants can be thought of as the generators
in the adjoint representation, $(T^{a}_{\rm adj})_{bc} \equiv \tf^{bac}$,
it is convenient to introduce a similar antisymmetry for
the fundamental generators,
\be
T^{\:\!a}_{\bar \jmath i} \equiv - T^a_{i \bar \jmath} ~~~~ \Leftrightarrow~~~~\tf^{cab} = - \tf^{bac} \,.
\label{signflip}
\ee
In formulas with suppressed indices
we denote the flipped generator by $\o{T}^a \equiv T^{\:\!a}_{\bar \imath j}$.

%%%%%%%%% FIGURE %%%%%%%%%%%%%%%
%%%%%%%%%%%%%%%%%%%%%%%%%%%%%%%%

\begin{figure}[t]
\centering
\includegraphics[scale=0.91,trim=0 0 0 0,clip=true]{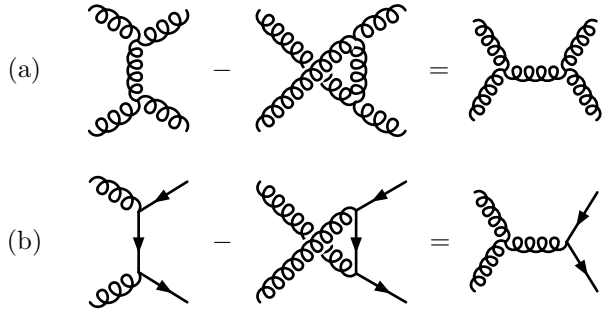}
\vspace{-3pt}
\caption{\small Color-algebra relations in the adjoint~(a)
                        and fundamental representation~(b).
The color-kinematics duality requires that
the kinematic numerators satisfy the corresponding kinematic-algebra relations,
which can be represented by the same graphs.}
\label{fig:fundjacobi}
\end{figure}

Color factors obey simple relations arising from
the Jacobi and commutation identities,
\begin{subequations} \begin{align}
   \tf^{dac} \tf^{cbe} - \tf^{dbc} \tf^{cae} & = \tf^{abc} \tf^{dce}
   \label{jacobi} \,, \\
   T^{a}_{i \bar \jmath} \, T^{b}_{j \bar k}   -
   T^{b}_{i \bar \jmath} \, T^{a}_{j \bar k} & = \tf^{abc} \, T^{c}_{i \bar k}
   \label{commutation} \,,
\end{align} \label{coloralgebra}%
\end{subequations}
depicted diagrammatically in \fig{fig:fundjacobi}.
They both imply color-algebraic relations of the schematic form
\be
   c_i-c_j=c_k
\ee
for the triplets of diagrams $(i,j,k)$
that differ only by the subdiagrams drawn in \fig{fig:fundjacobi},
but otherwise have common graph structure.
The interdependence among the color factors~$c_i$ means that
the corresponding kinematic coefficients~$n_i/D_i$ are in general not unique, as should be expected from the underlying gauge dependence of the numerators.
In \sec{sec:kinematicreview} we return to this kinematic numerator freedom. Before that, we review how to assemble the diagrams into gauge-invariant building blocks, \ie the primitive amplitudes.

%%%%%%%%% FIGURE %%%%%%%%%%%%%%%
%%%%%%%%%%%%%%%%%%%%%%%%%%%%%%%%

%%%%%%%%%%%%%%%%%%%%%%%%%%%%%%%%%%%%%%%%%%%%%%%%%%%%
\subsection{Color decomposition of tree amplitudes}
\label{sec:colorreview}
%%%%%%%%%%%%%%%%%%%%%%%%%%%%%%%%%%%%%%%%%%%%%%%%%%%%

A classic way to remove the relations among color factors is
to replace all structure constants by generators,
\be
   \tf^{abc} = \Tr\!\big( T^a T^b T^c \big ) - \Tr\!\big( T^b T^a T^c \big) \,,
\ee
and then eliminate all contracted adjoint indices using
the SU($N_c$) Fierz identity,
\be
   T^a_{i \bar \jmath} \:\! T^a_{k \bar l} =
      \delta_{i \bar l} \:\! \delta_{k \bar \jmath}
    - \frac{1}{N_c} \delta_{i \bar \jmath} \:\! \delta_{k \bar l} \,.
\label{ColorFierz}
\ee
This leads to a basis of linearly-independent color structures.
In the pure-gluon case it gives the familiar color-trace decomposition~\citeColorOrdering,
\be
    {\cal A}^{\text{tree}}_{n,0} = \!\!
      \sum_{\sigma \in S_{n-1}(\{2,\dots,n\})} \!\!\!
      \Tr\!\big( T^{a_1} T^{a_{\sigma(2)}} \dots T^{a_{\sigma(n)}} \big)
      A(1,\sigma(2),\dots,\sigma(n)) \,,
\label{GeneralGluonDecomposition}
\ee
where the sum is over $(n-1)!$ primitives because
the cyclic symmetry of the trace allows one to fix the first argument.\footnote{The
reversal symmetry $ A(1,2,\dots,n) = (-1)^n A(n,\dots,2,1) $
further reduces that count to $(n-1)!/2$.}
In a similar but much more cumbersome way, the color content of amplitudes
with $k$ quark-antiquark pairs and $(n-2k)$ gluons can be reduced~\cite{Cvitanovic:1980bu,Kosower:1988kh,Mangano:1988kk,Mangano:1990by,Ellis:2011cr,Ita:2011ar,Badger:2012pg,Schuster:2013aya,Reuschle:2013qna}
to color structures of the following type:
\be
   \frac{1}{N_c^{p}}
      \big( T^{a_{2k+1}}\!\dots T^{a_{l_1}} \big)_{i_1 \bar \alpha_1}
      \big( T^{a_{l_1+1}}\!\dots T^{a_{l_2}} \big)_{i_2 \bar \alpha_2} \dots
      \big( T^{a_{l_{k-1}+1}}\!\dots T^{a_n} \big)_{i_k \bar \alpha_k} \,,
\label{GeneralQuarkGluonDecomposition}
\ee
where $l_f \in \{ 2k+1, \dots,n \}$, $\bar \alpha_f \in \{ \bar \jmath_1, \dots, \bar \jmath_k \}$, and $(p+1)$ counts the number of disjoint cycles
in the permutation $\bar \alpha$.

An obvious feature of the above SU($N_c$) decomposition is that
it is specific to the gauge group.
A more interesting drawback is that it usually maps the color space to a basis
that is larger than the number of linearly independent color factors $c_i$.
Hence the resulting kinematic coefficients
-- the color-ordered amplitudes -- are not the minimal set of primitives.
In other words, they are not independent.
Indeed, the purely gluonic color-ordered amplitudes
can be reduced using the Kleiss-Kuijf (KK) relations~\cite{Kleiss:1988ne}.
By construction,
the number of the primitives independent under such color-algebra relations
must coincide with the number of linearly independent color factors.
The KK amplitude relations can be written as
\be
   A(1,\beta,2,\alpha) = (-1)^{|\beta|} \!\!\!
      \sum_{\sigma \in\:\!\alpha \shuffle \beta^T} \!\!
      A(1,2,\sigma) \,,
\label{KK}
\ee
where the sum runs over the shuffle product of the ordered sets
$\alpha$ and $\beta^T$, the latter being $\beta$ in reverse order.
This gives all partially ordered permutations that respect the element order of the two sets.
The KK relations let us fix the second argument in the pure gluon primitives,
and hence they reduce the basis to $(n-2)!$ elements -- the KK basis. 

%%%%%%%%% Figure %%%%%%%%%%%%%%%
\begin{figure}[t]
\centering
\includegraphics[scale=1.0,trim=0 0 0 0,clip=true]{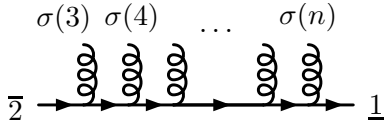}
\caption{\small Multi-peripheral cubic diagram for the color factors
         in formulas~\eqref{QuarkLineDecomposition} and~\eqref{DDM}.
         All permuted legs are gluons,
         while the horizontal line can be either a quark or a gluon line.}
\label{fig:MultiPeripheral}
\end{figure}
%%%%%%%%%%%%%%%%%%%%%%%%%%%%%%%%

An interesting exception to the redundancy
of the aforementioned decomposition algorithm is the case of the amplitude
with a single quark line~\cite{Kosower:1987ic,Mangano:1988kk},
\be
    {\cal A}^{\text{tree}}_{n,1} = \!\!
      \sum_{\sigma \in S_{n-2}(\{3,\dots,n\})} \!\!\!
      \big( T^{a_{\sigma(3)}} \dots T^{a_{\sigma(n)}}
      \big)_{\bar \jmath_2 i_1}
      A(\u{1},\o{2},\sigma(3),\dots,\sigma(n)) \,,
\label{QuarkLineDecomposition}
\ee
where we have denoted the quark-antiquark pair
by a bar below and above the labels, \ie as $\u{1}$ and $\o{2}$.
The sum is over the basis of $(n-2)!$ primitives, which are already
independent under color-algebra relations.
Moreover, the color factors in \eqn{QuarkLineDecomposition} are valid
for any gauge group, as can be guessed from the absence of explicit factors of $N_c$.

The reason for the nice properties of \eqn{QuarkLineDecomposition} is that this example happens to coincide
with another basis of color factors that have these properties more generally.
For $k<2$ the color factors in this basis correspond to the ``multi-peripheral'' graphs shown in \fig{fig:MultiPeripheral}. The above $k=1$ case can be easily mapped to the pure-gluon configuration
by replacing $T^a \rightarrow T^a_{\rm adj}$.
This gives the decomposition
of Del Duca, Dixon and Maltoni (DDM)~\cite{DelDuca:1999ha,DelDuca:1999rs}:
\be
    {\cal A}^{\text{tree}}_{n,0} = \!\!
      \sum_{\sigma \in S_{n-2}(\{3,\dots,n\})} \!\!
      \tf^{\,a_2 a_{\sigma(3)} b_1} \tf^{\,b_1 a_{\sigma(4)} b_2} \dots
%      \tf^{\,b_{n-4} a_{\sigma(n-1)} b_{n-3}}
      \tf^{\,b_{n-3} a_{\sigma(n)} a_1}
     \, A(1,2,\sigma(3),\dots,\sigma(n)) \,.
\label{DDM}
\ee
Note that it is a substantial improvement over
the trace decomposition~\eqref{GeneralGluonDecomposition},
since it avoids using $(n-2)^2(n-3)!$ primitives altogether.

The details become more involved when considering generalizations along the lines of the DDM decomposition to more than one quark line.  A basis of amplitude primitives for generic tree amplitudes in QCD, which have the same count as the number of independent color structures, was recently found by Melia in~\rcites{Melia:2013bta,Melia:2013epa}. However, it remained unknown what are the corresponding color coefficients in a decomposition using that basis.

In \sec{sec:coloralgebra} we discuss the Melia basis in detail, and give the complete amplitude color decomposition in terms of this basis. The new decomposition can be thought of as a natural generalization of the DDM decomposition to the case of $k$ quark-antiquark pairs, in analogy to how the Melia basis is the multi-quark generalization of the KK basis for amplitude primitives. Similarly to the multi-peripheral formulas~\eqref{QuarkLineDecomposition} and~\eqref{DDM}, the color coefficients of the new decomposition will be constructed from the cubic color factors~$c_i$ and will thus be valid for any gauge group (and any group representation for the quarks).

%%%%%%%%%%%%%%%%%%%%%%%%%%%%%%%%%%%%%%%%%%%%%%%%%%%%
\subsection{Color-kinematics duality}
\label{sec:kinematicreview}
%%%%%%%%%%%%%%%%%%%%%%%%%%%%%%%%%%%%%%%%%%%%%%%%%%%%

Let us return to the trivalent graph expansion~\eqref{BCJformYM}
of a gauge-theory amplitude, involving kinematic numerator factors $n_i$, color factors $c_i$ and denominators $D_i$. As already explained, the numerators are not uniquely defined, since, for instance, the quartic-vertex contact terms can be freely absorbed into the cubic graph numerators in more than one way. More generally, the ambiguity of the $n_i$ is a necessary consequence of the gauge dependence of individual Feynman diagrams, and hence the operation of shifting the numerators while leaving the amplitude invariant is called generalized gauge transformation~\cite{Bern:2008qj,Bern:2010ue}. 

It was observed by Bern, Carrasco and one of the current authors (BCJ)~\cite{Bern:2008qj}, that within the freedom of shifting the numerators there exists particularly nice choices, such that the resulting kinematic numerators $n_i$ obey the same general algebraic identities as the color factors $c_i$.
That is, there is a numerator relation for every
color Jacobi/commutation relation~\eqref{coloralgebra}
and a numerator sign flip for every color factor sign flip~\eqref{signflip}:
\begin{subequations} \begin{align}
      n_i - n_j = n_k ~~~&\Leftrightarrow~~~ c_i - c_j=c_k \,, 
\label{duality3} \\
      n_i \rightarrow -n_i  ~~~&\Leftrightarrow~~~ c_i \rightarrow -c_i \,.
\label{duality2}
\end{align} \label{duality}%
\end{subequations}
Amplitudes that satisfy these relations are said to exhibit color-kinematics duality. While both \eqns{duality3}{duality2} should be imposed, the latter is often omitted as the numerators inherit this antisymmetry from the cubic vertices of color-ordered Feynman rules.

The relations in \eqn{duality} define a kinematic algebra of numerators,
which is suggestive of an underlying kinematic Lie algebra. 
While not much is known about this Lie algebra,
which should be infinite-dimensional due to the continuous nature of momentum, in the special case of the self-dual sector of QCD the kinematic algebra is known to correspond to certain area-preserving diffeomorphisms~\cite{Monteiro:2011pc}.

A very useful aspect of the color-kinematics duality is that once the numerators satisfy \eqn{duality},
they can take the place of the color factors in \eqn{BCJformYM}, since they satisfy the same algebraic identities. Through this procedure one obtains a double copy amplitude,
\be
   {\cal M}^{\text{tree}}_{n,k} = i\;\!\Big(\frac{\kappa}{2}\Big)^{n-2}\!\!\!
      \sum_{\text{cubic graphs }\Gamma_i}^{\nu(n,k)} \!
      \frac{n_i \tilde{n}_i}{D_i} \,,
\label{BCJformGravity}
\ee
which describes scattering in a gravitational theory.
More precisely, the scattering amplitude~\eqref{BCJformGravity} will involve
$(n-2k)$ gravitons and $2k$ matter particles~\cite{Johansson:2014zca}. The tilde notation is introduced since the two copies of numerators may be not identical, they can differ by a generalized gauge transformation, or the states may differ on the two sides. 
More generally, the two sets of numerators in the
double copy construction do not have to belong to the same gauge theory~\cite{Bern:2008qj,Bern:2010ue},
producing a wide range of gravity theories with and without supersymmetry.
For example, pure Yang-Mills theory ``squares'' to
gravity coupled to a dilaton and anti-symmetric tensor.
Pure Einstein gravity can be obtained by removing these extra particles
via a ghost-like double-copy prescription
for massless quarks~\cite{Johansson:2014zca}. In contrast, a highly asymmetric double copy is needed for the amplitudes that couple Yang-Mills theory to gravity~\cite{Chiodaroli:2014xia}.

At tree level the double copy construction is known~\cite{Bern:2008qj,Bern:2010yg}
to be equivalent to the field-theory limit
of the Kawai-Lewellen-Tye (KLT) relations~\cite{Kawai:1985xq}
between open- and closed-string amplitudes.
However, the color-kinematics duality~\eqref{duality}
and the double copy~\eqref{BCJformGravity} can be argued to be deeper concepts, since they have straightforward generalizations to loop amplitudes~\cite{Bern:2010ue}. Amplitudes up to four loops have been constructed exhibiting the duality and double copy~\cite{Bern:2010ue,Carrasco:2011mn,Bern:2011rj,Bern:2012uf,Carrasco:2012ca,Boels:2013bi,Bjerrum-Bohr:2013iza,Bern:2013yya,Nohle:2013bfa,Bern:2013uka,Chiodaroli:2013upa,Johansson:2014zca,Bern:2014sna}. The implications of the color-kinematics duality have been used to derive a number of highly impressive results for string-theory amplitudes~\cite{BjerrumBohr:2009rd, Stieberger:2009hq,Mafra:2011nv,Mafra:2011nw,Mafra:2012kh,Broedel:2013tta,Stieberger:2014hba, Mafra:2014oia}, and more generally the duality has been studied using string-theory methods~\cite{Tye:2010dd,Mafra:2011kj,Barreiro:2013dpa,Ochirov:2013xba,Mafra:2015mja}. Recently the double-copy construction has been extended to classical Kerr-Schild solutions in general relativity and gauge theory~\cite{Monteiro:2014cda}.

In this paper the details of gravity amplitudes will not be discussed any further, since here we are instead interested in the aspect that the color-kinematics duality implies that primitive amplitudes obey
the Bern-Carrasco-Johansson (BCJ) relations~\cite{Bern:2008qj}. These relations constrain the primitive amplitudes
beyond the color-algebra basis discussed in the previous section.
In the pure-gluon case these kinematic-algebra relations
can be used to reduce the amplitude basis to $(n-3)!$ primitives.
The simplest family of BCJ relations is linear
in the generalized Mandelstam invariants:\footnote{These relations for gluon amplitudes
are sometimes called ``fundamental BCJ relations'' \cite{Feng:2010my}.
Here we avoid this terminology to prevent confusion
with the fundamental representation of the gauge group.}
\be
   \sum_{i=2}^{n-1} \Big( \sum_{j=2}^{i} s_{jn} \Big)\,
      A(1, 2, \ldots, i,n,i+1, \ldots, n-1) = 0 \,,
\label{fBCJ}
\ee
while the more general BCJ relations (found in~\rcite{Bern:2008qj})
can be derived~\cite{Feng:2010my} from \eqn{fBCJ} and its relabelings.

The BCJ amplitude relations motivated the discovery of the scattering equations and associated elegant string-like formulas for gauge and gravity tree amplitudes~\cite{Cachazo:2013iaa,
Cachazo:2013gna,Cachazo:2013hca,Cachazo:2013iea}. In the context of scattering equations the BCJ relations have been applied to a massless quark-antiquark pair~\cite{Weinzierl:2014ava} and a massive pair of scalars~\cite{Naculich:2014naa}, and three massive particles of different spins~\cite{Naculich:2015zha}.
In \sec{sec:kinematicalgebra}, we show that
the color-kinematics duality~\eqref{duality} holds for general QCD amplitudes, including an arbitrary number of massive quarks, and give an explicit solution to the resulting BCJ amplitude relations.

%%%%%%%%%%%%%%%%%%%%%%%%%%%%%%%%%%%%%%%%%%%%%%%%%%%%
\section{Color-algebra basis for quark-gluon amplitudes}
\label{sec:coloralgebra}
%%%%%%%%%%%%%%%%%%%%%%%%%%%%%%%%%%%%%%%%%%%%%%%%%%%%

In this section we analyze the color structure
of a general tree amplitude in QCD:
we discuss the Melia basis~\cite{Melia:2013epa} of $(n-2)!/k!$ color-ordered primitives and find their color coefficients.
This will constitute a new color decomposition for QCD amplitudes.
Though the results are applicable to arbitrary matter particles
in any gauge-group representation, we refer to them as quarks.
Similarly, the gluon can in principle be replaced by
any adjoint particle since only group-theoretic properties are used in this section.

We start with the case with no gluons, \ie the multi-quark amplitudes,
with all $k$ quark lines having different flavors.
In \rcite{Melia:2013bta} Melia considered such amplitudes with quarks
in the adjoint representation and found a basis of primitives
independent under all color-algebra amplitude relations.
The latter are simply the familiar KK relations~\eqref{KK}
projected onto the multi-quark case
by setting to zero all the diagrams and primitives with crossed quark flavor lines (crossed with respect to a planar color ordering).

This is possible because the KK relations can be deduced~\cite{Bern:2008qj}
from the fact that color-ordered primitive amplitudes
can be expanded in terms of antisymmetric cubic vertices,
as is manifest in the adjoint representation, and can be imposed in the fundamental case (see \fig{fig:colorvertices}).
Under the projection that removes crossed fermion line diagrams, the antisymmetry of the vertices is maintained (since zero is antisymmetric) thus all the fermion-case KK relations are inherited from the pure-gluon case unscathed,
though some of them are reduced to redundant equations,
or even to trivial $0 = 0$ equations.
This implies that the fermion-case basis of primitives has to be a subset of the original basis of $(n-2)!$ primitives.
In fact, the $(n-2)!$ basis decomposes into $k!=(n/2)!$ cases
of inequivalent permutations of the quarks
(maintaining the antiquark ordering),
out of which only one has no flavor crossings. Hence,
the basis surviving the projection is reduced in size by a factor $1/(n/2)!$.
Indeed, the Melia basis for pure-fermion amplitudes has dimension $(n-2)!/(n/2)!$~\cite{Melia:2013bta}, and in the general mixed quark-gluon case the dimension is $(n-2)!/k!$~\cite{Melia:2013epa}.

The fact that the basis of primitives were derived in~\rcites{Melia:2013bta,Melia:2013epa} for the case of adjoint particles and color-ordered planar amplitudes is not a problem. The same planar amplitudes are also a basis of primitives in the mixed adjoint-fundamental case that we encounter in QCD. We illustrate this through examples, and then give the general color decomposition for QCD.

%%%%%%%%%%%%%%%%%%%%%%%%%%%%%%%%%%%%%%%%%%%%%%%%%%%%
\subsection{Pure-quark example: $n=6$, $k=3$}
\label{sec:multiquark}
%%%%%%%%%%%%%%%%%%%%%%%%%%%%%%%%%%%%%%%%%%%%%%%%%%%%

Let us now explain the details of the Melia basis
using an instructive pure-quark example.

The basis of primitive amplitudes
for $k=n/2$ quark-antiquark pairs
is given by~\cite{Melia:2013bta}\footnote{The precise basis
used in \rcite{Melia:2013bta} is slightly different:
$ \big\{ A(\o{2},\u{1},\sigma) \,\big|\,\sigma \in \text{Dyck}_{k-1} \big\} $
in our notation.}
\be
   \Big\{ A(\u{1},\o{2},\sigma) ~\big|~
          \sigma \in \text{Dyck}_{k-1} \Big\} \,,
\label{MeliaBasis}
\ee
where the last $(n-2)$ arguments must form a valid Dyck word
out of the quarks and antiquark labels.\footnote{Recall that
we mark quarks and antiquarks with underscores and overscores,
respectively.}
These words are defined as strings of X and Y letters, $(k-1)$ of each,
such that the number of X's preceding each Y
is greater than the number of preceding Y's.
A more illuminating representation is obtained by realizing that
Dyck words precisely correspond to well-formed brackets,
with X's playing the role of opening brackets
and Y's being closing brackets.
For $n=6$ there are two such words, XYXY and XXYY,
equivalent to the brackets $\{\}\{\}$ and $\{\{\}\}$, respectively.
An example of an invalid Dyck word is XYYX,
and it translates to the ill-formed brackets $\{\}\}\{$.

Of course, the brackets themselves are not sufficient to specify
the particle labels, instead they only specify the particle type.
The X's or opening brackets can be identified with quarks
and Y's or closing brackets with antiquarks (or, more precisely,
the fundamental and anti-fundamental representation, respectively).
To form the Dyck words relevant for \eqn{MeliaBasis},
each valid bracket combination needs to be populated by particle labels.
To be specific, if we have quark flavor lines $\u{3} \leftarrow \o{4}$
and $\u{5} \leftarrow  \o{6}$,
then there are exactly two label assignments per bracket combination
that leave the flavor lines uncrossed:
\begin{subequations} \begin{align}
   \label{Dyck6pointPlanar}
 & \text{XYXY}~~\Rightarrow~~(\u{3},\o{4},\u{5},\o{6}) ,\,
                           (\u{5},\o{6},\u{3},\o{4})
                         ~\Leftrightarrow~ \{3\:4\}\{5\:6\} ,\,
                           \{5\:6\}\{3\:4\} \,, \\
   \label{Dyck6pointNonplanar}
 & \text{XXYY}~~\Rightarrow~~(\u{3},\u{5},\o{6},\o{4}) ,\,
                           (\u{5},\u{3},\o{4},\o{6})
                         ~\Leftrightarrow~ \big\{3\{5\:6\}4\big\} ,\,
                           \big\{5\{3\:4\}6\big\} \,.
\end{align} \label{Dyck6point}%
\end{subequations}
These four valid Dyck words are written out using two different notations:
the bar notation used in \eqn{QuarkLineDecomposition}
and the bracket notation introduced above.
As can be seen, the two notations convey the same information,
since each downstairs and upstairs bar can be respectively identified
with an opening and a closing bracket.
In this paper we use both notations interchangeably according to convenience.
A very important aspect of either notation is that each opening bracket corresponds to a unique closing bracket, and hence each downstairs bar corresponds to a unique upstairs bar. These pairs of states are precisely the quark-antiquark pairs of the same flavor. Thus the notation conveniently combines the information about the gauge-group representation and flavor.
In other words, no separate notation is needed for specifying the flavor of the primitive amplitudes.

%%%%%%%%% FIGURE %%%%%%%%%%%%%%%
\begin{figure}[t]
\centering
\includegraphics[scale=1.0,trim=0 0 0 0,clip=true]{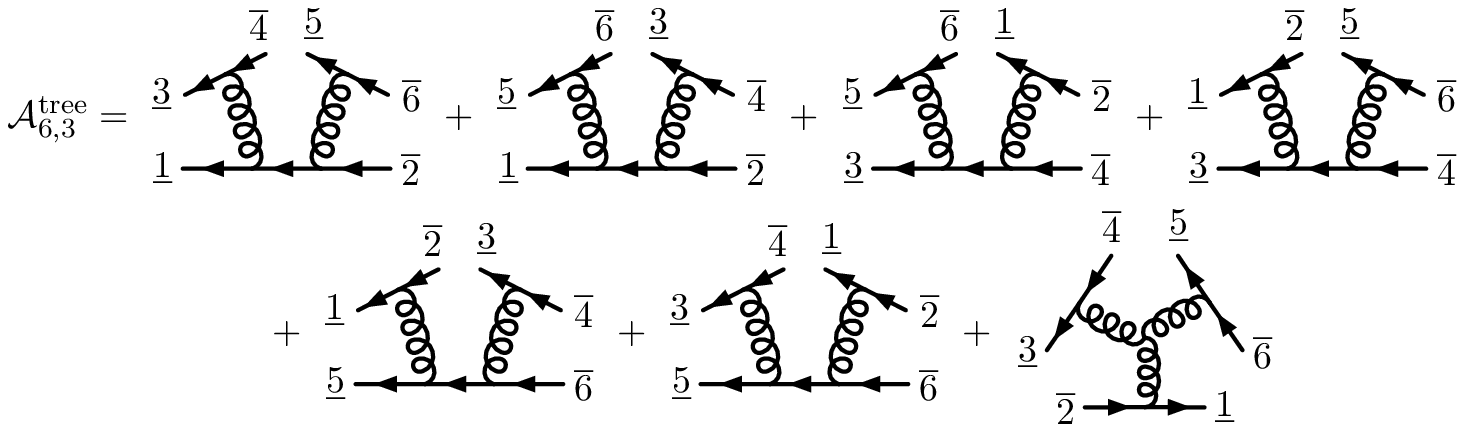}
\caption{\small Feynman diagrams for the six-quark amplitude
         ${\cal A}^{\text{tree}}_{6,3}(\u{1},\o{2},\u{3},\o{4},\u{5},\o{6})$.}
\label{fig:qqQQqq}
\end{figure}
%%%%%%%%%%%%%%%%%%%%%%%%%%%%%%%%

We have thus arrived at the Melia basis for $n=6$, $k=3$,
which contains four primitives:
\be
A(\u{1},\o{2},\u{3},\o{4},\u{5},\o{6}) \,,~
A(\u{1},\o{2},\u{5},\o{6},\u{3},\o{4}) \,,~
A(\u{1},\o{2},\u{3},\u{5},\o{6},\o{4}) \;\text{ and }
A(\u{1},\o{2},\u{5},\u{3},\o{4},\o{6}) \,.
\label{Melia6point}
\ee
Now we will determine the color factors corresponding to the primitives and that are appropriate for quarks
in the fundamental representation.
Recall that the full amplitude can be written as
\be
   {\cal A}_{6,3}^{\text{tree}} =
      \frac{c_1 n_1}{D_1} + \frac{c_2 n_2}{D_2}
    + \frac{c_3 n_3}{D_3} + \frac{c_4 n_4}{D_4}
    + \frac{c_5 n_5}{D_5} + \frac{c_6 n_6}{D_6}
    + \frac{c_7 n_7}{D_7} \,,
\label{AqqQQqq}
\ee
where each term corresponds to a cubic Feynman diagram in \fig{fig:qqQQqq}.
Their color factors
\begin{subequations} \begin{align}
 & c_1 = T^a_{i_1 \bar \jmath} T^b_{j\,\bar \imath_2}
         T^a_{i_3 \bar \imath_4} T^b_{i_5 \bar \imath_6} \,, \qquad \qquad
   c_2 = T^b_{i_1 \bar \jmath} T^a_{j\,\bar \imath_2}
         T^a_{i_3 \bar \imath_4} T^b_{i_5 \bar \imath_6} \,, \\
 & c_3 = T^a_{i_3 \bar \jmath} T^b_{j\,\bar \imath_4}
         T^a_{i_5 \bar \imath_6} T^b_{i_1 \bar \imath_2} \,, \qquad \qquad
   c_4 = T^b_{i_3 \bar \jmath} T^a_{j\,\bar \imath_4}
         T^a_{i_5 \bar \imath_6} T^b_{i_1 \bar \imath_2} \,, \\
 & c_5 = T^a_{i_5 \bar \jmath} T^b_{j\,\bar \imath_6}
         T^a_{i_1 \bar \imath_2} T^b_{i_3 \bar \imath_4} \,, \qquad \qquad
   c_6 = T^b_{i_5 \bar \jmath} T^a_{j\,\bar \imath_6}
         T^a_{i_1 \bar \imath_2} T^b_{i_3 \bar \imath_4} \,, \\
 & \hspace{70pt}
   c_7 = -\tilde{f}^{abc}\;\!T^a_{i_1 \bar \imath_2}
         T^b_{i_3 \bar \imath_4} T^c_{i_5 \bar \imath_6} \,,
\end{align} \label{color6point}%
\end{subequations}
can be read off according the rules in \fig{fig:colorvertices}.
For completeness and later use, we give the kinematic content
of three representative diagrams:
\begin{subequations} \begin{align}
 & n_1 = -\frac{i}{4}
         (\bar{u}_1 \gamma^{\mu} ({\not}k_{1,3,4}\!+\!m_1) \gamma^{\nu} v_2)
         (\bar{u}_3 \gamma_{\mu} v_4) (\bar{u}_5 \gamma_{\nu} v_6) \,,
   \qquad
   D_1 = (s_{1,3,4} - m_1^2) s_{34} s_{56} \,, \\
 & n_2 = -\frac{i}{4}
         (\bar{u}_1 \gamma^{\nu} ({\not}k_{1,5,6}\!+\!m_1) \gamma^{\mu} v_2)
         (\bar{u}_3 \gamma_{\mu} v_4) (\bar{u}_5 \gamma_{\nu} v_6) \,,
   \qquad
   D_2 = (s_{1,5,6} - m_1^2) s_{34} s_{56} \,, \\
 & n_7 = -\frac{i}{4}
         (\bar{u}_1 \gamma^{\mu} v_2) (\bar{u}_3 \gamma_{\mu} v_4)
         (\bar{u}_5 ({\not}k_{1,2} - {\not}k_{3,4}) v_6) + \text{cyclic} \,,
   \qquad \qquad ~\:
   D_7 = s_{12} s_{34} s_{56} \,,
\end{align} \label{numerators6point}%
\end{subequations}
where we allow for arbitrary spins.
All momenta are outgoing and the notation used is
\beal
   k_{i,j,\dots,l} & = k_i + k_j + \dots + k_l \,, \\
   s_{i,j,\dots,l} & = (k_i + k_j + \dots + k_l)^2 \,, \\
   s_{ij} & = (k_i+k_j)^2  \,.
\label{KinematicNotation}
\eeal

It is easy to see that the seven color factors~\eqref{color6point}
obey three commutation relations:
\beal
   c_1 - c_2 = -c_7 \,, \qquad \qquad
   c_3 - c_4 = -c_7 \,, \qquad \qquad
   c_5 - c_6 = -c_7 \,.
\label{crelations6point}
\eeal
Thus the color space of ${\cal A}_{6,3}^{\text{tree}}$ is four-dimensional,
which is consistent with the length of the Melia basis~\eqref{Melia6point}.
The primitive amplitudes therein can also be diagrammatically expanded
according to the color-ordered Feynman rules~\eqref{kinematicvertices}
\beal
 & A_{\u{1}\o{2}\u{3}\o{4}\u{5}\o{6}} =
      \frac{n_2}{D_2} + \frac{n_4}{D_4}
    + \frac{n_6}{D_6} + \frac{n_7}{D_7} \,, \qquad
   A_{\u{1}\o{2}\u{3}\u{5}\o{6}\o{4}} =
    - \frac{n_3}{D_3} - \frac{n_4}{D_4} \,, \\
 & A_{\u{1}\o{2}\u{5}\o{6}\u{3}\o{4}} =
      \frac{n_1}{D_1} + \frac{n_3}{D_3}
    + \frac{n_5}{D_5} - \frac{n_7}{D_7} \,, \qquad
   A_{\u{1}\o{2}\u{5}\u{3}\o{4}\o{6}} =
    - \frac{n_5}{D_5} - \frac{n_6}{D_6} \,,
\label{primitives6point}
\eeal
where for brevity we have written the amplitude arguments as subscripts. 

Now we wish to color-decompose the amplitude~\eqref{AqqQQqq} such that
we get exactly the Melia basis primitives as kinematic coefficients.
We observe that we can massage it
into a combination of the expressions in \eqn{primitives6point}
by using the commutation relations~\eqref{crelations6point}.
Indeed, if we eliminate the color factors $c_3$, $c_6$ and $c_7$,
we land on the following decomposition:
\begin{align}
   {\cal A}_{6,3}^{\text{tree}} & =
      \frac{c_1 n_1}{D_1} + \frac{c_2 n_2}{D_2}
    + \frac{(c_1-c_2+c_4) n_3}{D_3} + \frac{c_4 n_4}{D_4}
    + \frac{c_5 n_5}{D_5} + \frac{(c_2-c_1+c_5) n_6}{D_6}
    + \frac{(c_2-c_1) n_7}{D_7}  \nn \\
  & \!\begin{aligned}
  & = c_2\,A_{\u{1}\o{2}\u{3}\o{4}\u{5}\o{6}}
    + c_1\,A_{\u{1}\o{2}\u{5}\o{6}\u{3}\o{4}}
    + (c_2 - c_4)\,A_{\u{1}\o{2}\u{3}\u{5}\o{6}\o{4}}
    + (c_1 - c_5)\,A_{\u{1}\o{2}\u{5}\u{3}\o{4}\o{6}} \\
  & \equiv
      C_{\u{1}\o{2}\u{3}\o{4}\u{5}\o{6}}\,A_{\u{1}\o{2}\u{3}\o{4}\u{5}\o{6}}
    + C_{\u{1}\o{2}\u{5}\o{6}\u{3}\o{4}}\,A_{\u{1}\o{2}\u{5}\o{6}\u{3}\o{4}}
    + C_{\u{1}\o{2}\u{3}\u{5}\o{6}\o{4}}\,A_{\u{1}\o{2}\u{3}\u{5}\o{6}\o{4}} 
    + C_{\u{1}\o{2}\u{5}\u{3}\o{4}\o{6}}\,A_{\u{1}\o{2}\u{5}\u{3}\o{4}\o{6}} \,.
   \end{aligned}
\label{tree6point}
\end{align}
The precise expressions for $C_{\u{1}\o{2}\dots}$'s, 
the color coefficients of the primitives, are subject
to the commutation identities~\eqref{crelations6point}. 
Our choice here is the one that will be generalized in \sec{sec:colorfactors}.
Some traits of the general pattern can be seen from
the following rendition of the color coefficients:
\beal \hspace{-10pt}
 & C_{\u{1}\o{2}\u{3}\o{4}\u{5}\o{6}} =
      \parbox{110pt}{
      \begin{fmffile}{c123456} \fmfframe(10,3)(0,3){
      \fmfsettings \small
      \begin{fmfgraph*}(90,25)
      \fmfbottom{q2,,,q1}
      \fmftop{q3,,,q6}
      \fmf{phantom,tension=100}{q3,d3,q4}
      \fmf{phantom,tension=70}{q4,q5}
      \fmf{phantom,tension=100}{q5,d6,q6}
      \fmf{phantom,tension=100}{q2,b3,b4}
      \fmf{phantom,tension=70}{b4,b5}
      \fmf{phantom,tension=100}{b5,b6,q1}
      \fmf{plain_arrow,tension=0}{q2,b3,b6,q1}
      \fmf{plain_arrow,tension=0}{q4,d3,q3}
      \fmf{plain_arrow,tension=0}{q6,d6,q5}
      \fmf{curly,tension=0}{b3,d3}
      \fmf{curly,tension=0}{b6,d6}
      \fmfv{label=$\u{1}$,label.angle=0,label.dist=2.5pt}{q1}
      \fmfv{label=$\o{2}$,label.angle=180,label.dist=2.5pt}{q2}
      \fmfv{label=$\u{3}$,label.angle=180,label.dist=2.5pt}{q3}
      \fmfv{label=$\u{5}$,label.angle=180,label.dist=2.5pt}{q5}
      \fmfv{label=$\o{4}$,label.angle=0,label.dist=2.5pt}{q4}
      \fmfv{label=$\o{6}$,label.angle=0,label.dist=2.5pt}{q6}
      \end{fmfgraph*} }
      \end{fmffile}
      } \,, \quad
   C_{\u{1}\o{2}\u{3}\u{5}\o{6}\o{4}} =
      \parbox{80pt}{
      \begin{fmffile}{c2c123564} \fmfframe(10,3)(0,3){
      \fmfsettings \small
      \begin{fmfgraph*}(60,50)
      \fmfbottom{q2,,,q1}
      \fmftop{l5,,,q6}
      \fmfleft{q3}
      \fmfright{q4}
      \fmf{phantom,tension=100}{l5,q5,u4,q6}
      \fmf{phantom,tension=100}{q3,d3,d4,q4}
      \fmf{phantom,tension=100}{q2,b3,b4,q1}
      \fmf{plain_arrow,tension=0}{q2,b3,b4,q1}
      \fmf{plain_arrow,tension=0}{q4,d3,q3}
      \fmf{plain_arrow,tension=0}{q4,d4}
      \fmf{plain_arrow,tension=0}{q6,u4,q5}
      \fmf{curly,tension=0}{b3,d3}
      \fmf{curly,tension=0,rubout=3}{b4,u4}
      \fmfv{label=$\u{1}$,label.angle=0,label.dist=2.5pt}{q1}
      \fmfv{label=$\o{2}$,label.angle=180,label.dist=2.5pt}{q2}
      \fmfv{label=$\u{3}$,label.angle=180,label.dist=2.5pt}{q3}
      \fmfv{label=$\u{5}$,label.angle=180,label.dist=2.5pt}{q5}
      \fmfv{label=$\o{4}$,label.angle=0,label.dist=2.5pt}{q4}
      \fmfv{label=$\o{6}$,label.angle=0,label.dist=2.5pt}{q6}
      \end{fmfgraph*} }
      \end{fmffile}
      }
      +
      \parbox{80pt}{
      \begin{fmffile}{c4c123564} \fmfframe(10,3)(0,3){
      \fmfsettings \small
      \begin{fmfgraph*}(60,50)
      \fmfbottom{q2,,,q1}
      \fmftop{l5,,,q6}
      \fmfleft{q3}
      \fmfright{q4}
      \fmf{phantom,tension=100}{l5,q5,u4,q6}
      \fmf{phantom,tension=100}{q3,d3,d4,q4}
      \fmf{phantom,tension=100}{q2,b3,b4,q1}
      \fmf{plain_arrow,tension=0}{q2,b3,q1}
      \fmf{plain_arrow,tension=0}{q4,d4,d3,q3}
      \fmf{plain_arrow,tension=0}{q6,u4,q5}
      \fmf{curly,tension=0}{b3,d3}
      \fmf{curly,tension=0}{d4,u4}
      \fmfv{label=$\u{1}$,label.angle=0,label.dist=2.5pt}{q1}
      \fmfv{label=$\o{2}$,label.angle=180,label.dist=2.5pt}{q2}
      \fmfv{label=$\u{3}$,label.angle=180,label.dist=2.5pt}{q3}
      \fmfv{label=$\u{5}$,label.angle=180,label.dist=2.5pt}{q5}
      \fmfv{label=$\o{4}$,label.angle=0,label.dist=2.5pt}{q4}
      \fmfv{label=$\o{6}$,label.angle=0,label.dist=2.5pt}{q6}
      \end{fmfgraph*} }
      \end{fmffile}
      } \,, \hspace{-10pt} \\ \hspace{-10pt}
 & C_{\u{1}\o{2}\u{5}\o{6}\u{3}\o{4}} =
      \parbox{110pt}{
      \begin{fmffile}{c125634} \fmfframe(10,3)(0,3){
      \fmfsettings \small
      \begin{fmfgraph*}(90,25)
      \fmfbottom{q2,,,q1}
      \fmftop{q3,,,q6}
      \fmf{phantom,tension=100}{q3,d3,q4}
      \fmf{phantom,tension=70}{q4,q5}
      \fmf{phantom,tension=100}{q5,d6,q6}
      \fmf{phantom,tension=100}{q2,b3,b4}
      \fmf{phantom,tension=70}{b4,b5}
      \fmf{phantom,tension=100}{b5,b6,q1}
      \fmf{plain_arrow,tension=0}{q2,b3,b6,q1}
      \fmf{plain_arrow,tension=0}{q4,d3,q3}
      \fmf{plain_arrow,tension=0}{q6,d6,q5}
      \fmf{curly,tension=0}{b3,d3}
      \fmf{curly,tension=0}{b6,d6}
      \fmfv{label=$\u{1}$,label.angle=0,label.dist=2.5pt}{q1}
      \fmfv{label=$\o{2}$,label.angle=180,label.dist=2.5pt}{q2}
      \fmfv{label=$\u{5}$,label.angle=180,label.dist=2.5pt}{q3}
      \fmfv{label=$\u{3}$,label.angle=180,label.dist=2.5pt}{q5}
      \fmfv{label=$\o{6}$,label.angle=0,label.dist=2.5pt}{q4}
      \fmfv{label=$\o{4}$,label.angle=0,label.dist=2.5pt}{q6}
      \end{fmfgraph*} }
      \end{fmffile}
      } \,, \quad
   C_{\u{1}\o{2}\u{5}\u{3}\o{4}\o{6}} =
      \parbox{80pt}{
      \begin{fmffile}{c2c125346} \fmfframe(10,3)(0,3){
      \fmfsettings \small
      \begin{fmfgraph*}(60,50)
      \fmfbottom{q2,,,q1}
      \fmftop{l5,,,q6}
      \fmfleft{q3}
      \fmfright{q4}
      \fmf{phantom,tension=100}{l5,q5,u4,q6}
      \fmf{phantom,tension=100}{q3,d3,d4,q4}
      \fmf{phantom,tension=100}{q2,b3,b4,q1}
      \fmf{plain_arrow,tension=0}{q2,b3,b4,q1}
      \fmf{plain_arrow,tension=0}{q4,d3,q3}
      \fmf{plain_arrow,tension=0}{q4,d4}
      \fmf{plain_arrow,tension=0}{q6,u4,q5}
      \fmf{curly,tension=0}{b3,d3}
      \fmf{curly,tension=0,rubout=3}{b4,u4}
      \fmfv{label=$\u{1}$,label.angle=0,label.dist=2.5pt}{q1}
      \fmfv{label=$\o{2}$,label.angle=180,label.dist=2.5pt}{q2}
      \fmfv{label=$\u{5}$,label.angle=180,label.dist=2.5pt}{q3}
      \fmfv{label=$\u{3}$,label.angle=180,label.dist=2.5pt}{q5}
      \fmfv{label=$\o{6}$,label.angle=0,label.dist=2.5pt}{q4}
      \fmfv{label=$\o{4}$,label.angle=0,label.dist=2.5pt}{q6}
      \end{fmfgraph*} }
      \end{fmffile}
      }
      +
      \parbox{80pt}{
      \begin{fmffile}{c6c125346} \fmfframe(10,3)(0,3){
      \fmfsettings \small
      \begin{fmfgraph*}(60,50)
      \fmfbottom{q2,,,q1}
      \fmftop{l5,,,q6}
      \fmfleft{q3}
      \fmfright{q4}
      \fmf{phantom,tension=100}{l5,q5,u4,q6}
      \fmf{phantom,tension=100}{q3,d3,d4,q4}
      \fmf{phantom,tension=100}{q2,b3,b4,q1}
      \fmf{plain_arrow,tension=0}{q2,b3,q1}
      \fmf{plain_arrow,tension=0}{q4,d4,d3,q3}
      \fmf{plain_arrow,tension=0}{q6,u4,q5}
      \fmf{curly,tension=0}{b3,d3}
      \fmf{curly,tension=0}{d4,u4}
      \fmfv{label=$\u{1}$,label.angle=0,label.dist=2.5pt}{q1}
      \fmfv{label=$\o{2}$,label.angle=180,label.dist=2.5pt}{q2}
      \fmfv{label=$\u{5}$,label.angle=180,label.dist=2.5pt}{q3}
      \fmfv{label=$\u{3}$,label.angle=180,label.dist=2.5pt}{q5}
      \fmfv{label=$\o{6}$,label.angle=0,label.dist=2.5pt}{q4}
      \fmfv{label=$\o{4}$,label.angle=0,label.dist=2.5pt}{q6}
      \end{fmfgraph*} }
      \end{fmffile}
      } \,, \hspace{-10pt}
\label{cprimitives6point}
\eeal
where we choose to draw some of the color diagrams in a nonplanar fashion
in order to preserve the cyclic ordering dictated by the primitive amplitudes.
In other words, although the color-ordered primitives~\eqref{primitives6point}
are composed only of planar Feynman diagrams,
their color coefficients~\eqref{cprimitives6point}
contain both planar and nonplanar color diagrams.
More than that, we observe that this non-planarity is related to
the bracket ``nestedness'' for configurations $\big\{3\{5\:6\}4\big\}$
and $\big\{5\{3\:4\}6\big\}$,
as will be increasingly clear for higher-multiplicity examples.

%%%%%%%%%%%%%%%%%%%%%%%%%%%%%%%%%%%%%%%%%%%%%%%%%%%%
\subsection{Quark-gluon example: $n=5$, $k=2$}
\label{sec:5pointcolor}
%%%%%%%%%%%%%%%%%%%%%%%%%%%%%%%%%%%%%%%%%%%%%%%%%%%%

%%%%%%%%% FIGURE %%%%%%%%%%%%%%%
\begin{figure}[t]
\centering
\includegraphics[scale=1.0,trim=0 0 0 0,clip=true]{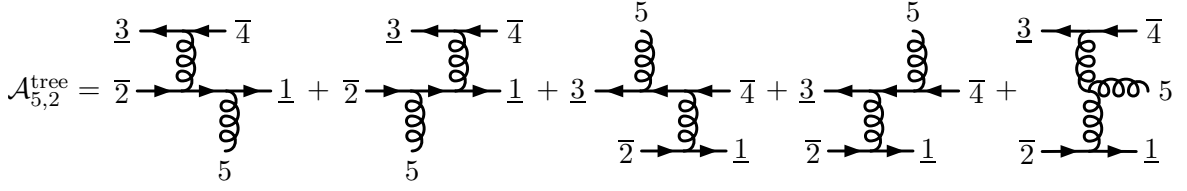}
\caption{\small Feynman diagrams for the four-quark one-gluon amplitude
         ${\cal A}^{\text{tree}}_{5,2}(\u{1},\o{2},\u{3},\o{4},5)$.}
\label{fig:qqQQg}
\end{figure}
%%%%%%%%%%%%%%%%%%%%%%%%%%%%%%%%

Now let us replace the last quark pair by a single gluon,
while four remaining quarks have the same distinct flavors as before.
The quark masses are then $m_1=m_2$, $m_3=m_4$.
For completeness and further use,
we give full color and kinematic content
for each of the five Feynman diagrams shown in \fig{fig:qqQQg}:
{\small
\begin{subequations} \begin{align}
 & c_1 = T_{i_1 \bar \jmath}^{a_5} T_{j\,\bar \imath_2}^b
         T_{i_3 \bar \imath_4}^b \,, \quad~
   n_1 = \frac{i}{2\sqrt{2}} \,
        ( \bar{u}_1 {\not}\varepsilon_5
                   ({\not}k_{1,5}\!+\!m_1) \gamma^\mu v_2 )
        ( \bar{u}_3 \gamma_\mu v_4 ) \,, \quad~
   D_1 = (s_{15}\!-\!m_1^2) s_{34} \,, \\
 & c_2 = T_{i_1 \bar \jmath}^b T_{j\,\bar \imath_2}^{a_5}
         T_{i_3 \bar \imath_4}^b \,, \quad~
   n_2 = \frac{-i}{2\sqrt{2}} \,
        ( \bar{u}_1 \gamma^\mu ({\not}k_{2,5}\!-\!m_2)
                                {\not}\varepsilon_5 v_2 )
        ( \bar{u}_3 \gamma_\mu v_4 ) \,, \quad~
   D_2 = (s_{25}\!-\!m_2^2) s_{34} \,, \\
 & c_3 = T_{i_1 \bar \imath_2}^b
         T_{i_3 \bar \jmath}^{a_5} T_{j\,\bar \imath_4}^b \,, \quad~
   n_3 = \frac{i}{2\sqrt{2}} \,
        ( \bar{u}_1 \gamma_\mu v_2 )
        ( \bar{u}_3 {\not}\varepsilon_5
                   ({\not}k_{3,5}\!+\!m_3) \gamma^\mu v_4 ) \,, \quad~
   D_3 = s_{12} (s_{35}\!-\!m_3^2) \,, \\
 & c_4 = T_{i_1 \bar \imath_2}^b
         T_{i_3 \bar \jmath}^b T_{j\,\bar \imath_4}^{a_4} \,, \quad~
   n_4 = \frac{-i}{2\sqrt{2}} \,
         ( \bar{u}_1 \gamma_\mu v_2 )
        ( \bar{u}_3 \gamma^\mu ({\not}k_{4,5}\!-\!m_4)
                                {\not}\varepsilon_5 v_4 ) \,, \quad~
   D_4 = s_{12} (s_{45}\!-\!m_4^2) \,, \\
 & c_5 = \tilde{f}^{a_5 b\,c}\;\! T_{i_1 \bar \imath_2}^b
                                  T_{i_3 \bar \imath_4}^c \,, ~
   n_5 = \frac{i}{\sqrt{2}}
          \Big( ( \bar{u}_1 {\not}\varepsilon_5 v_2 )
                ( \bar{u}_3 {\not}k_5 v_4 )
              - ( \bar{u}_1 {\not}k_5 v_2 )
                ( \bar{u}_3 {\not}\varepsilon_5 v_4 ) \\
& ~~~~~~~~~~~~~~~~~~~~~~~~~~~~~~~~~~~~~~~~~~~~~~~~~~~~\:
              -\, ( \bar{u}_1 \gamma^\mu v_2 )
                  ( \bar{u}_3 \gamma_\mu v_4 )
                  (k_{12}\!\cdot\!\varepsilon_5) \Big) \,, \quad
   D_5 = s_{12} s_{34} \,. \nn
\end{align} \label{numerators5point}%
\end{subequations}}
The five color factors obey two commutation relations:
\beal
   c_1 - c_2 = -c_5 \,, \qquad \qquad
   c_3 - c_4 =  c_5 \,.
\label{crelations5point}
\eeal
implying a color-algebra basis of three primitive amplitudes.
A symmetric color decomposition is obtained if we eliminate $c_1$ and $c_4$:
\beal
   {\cal A}_{5,2}^\text{tree} = \sum_{i=1}^5 \frac{c_i n_i}{D_i}
      & = c_2 \left( \frac{n_1}{D_1} + \frac{n_2}{D_2} \right)
        + c_3 \left( \frac{n_3}{D_3} + \frac{n_4}{D_4} \right)
        + c_5 \left(-\frac{n_1}{D_1} - \frac{n_4}{D_4}
                                     + \frac{n_5}{D_5} \right) \\
      & \equiv c_2 A_{\u{1} 5 \o{2}\u{3}\o{4}}
             + c_3 A_{\u{1}\o{2}\u{3} 5 \o{4}}
             + c_5 A_{\u{1}\o{2}\u{3}\o{4} 5} \,,
\label{tree5symmetric}
\eeal
where in the last line we identified the correct color-ordered amplitudes,
corresponding to the independent color factors $c_2$, $c_3$ and $c_5$
(similar decompositions are found in \rcite{Melia:2014oza}).
However, this is not yet the Melia basis~\cite{Melia:2013epa}.
To obtain it, we use the KK relation~\eqref{KK} to move leg $\o{2}$
next to leg $\u{1}$:
\be
   A(\u{1},\underbrace{5,}_\beta\o{2},\underbrace{\u{3},\o{4}}_\alpha\;\!)
      = - A(\u{1},\o{2},5,\u{3},\o{4}) - A(\u{1},\o{2},\u{3},5,\o{4})
                                       - A(\u{1},\o{2},\u{3},\o{4},5) \,.
\label{KK5}
\ee
Thus we obtain a new decomposition
\beal
   {\cal A}_{5,2}^\text{tree} \, &= \, - c_2 A_{\u{1}\o{2} 5 \u{3}\o{4}}
                                       - c_1 A_{\u{1}\o{2}\u{3}\o{4} 5}
                                 +(-c_1+c_4) A_{\u{1}\o{2}\u{3} 5 \o{4}} \\
   & \equiv \, C_{\u{1}\o{2} 5 \u{3}\o{4}}\,A_{\u{1}\o{2} 5 \u{3}\o{4}}
             + C_{\u{1}\o{2}\u{3}\o{4} 5}\, A_{\u{1}\o{2}\u{3}\o{4} 5}
             + C_{\u{1}\o{2}\u{3} 5 \o{4}}\,A_{\u{1}\o{2}\u{3} 5 \o{4}} \,,
\label{tree5point}
\eeal
in terms of three primitives
\be
   A_{\u{1}\o{2} 5 \u{3}\o{4}} =
    - \frac{n_2}{D_2} - \frac{n_3}{D_3} - \frac{n_5}{D_5} \,, \quad
   A_{\u{1}\o{2}\u{3}\o{4} 5} =
    - \frac{n_1}{D_1} - \frac{n_4}{D_4} + \frac{n_5}{D_5} \,, \quad
   A_{\u{1}\o{2}\u{3} 5 \o{4}} = \frac{n_3}{D_3} + \frac{n_4}{D_4} \,.
\label{primitives5point}
\ee
Note that they correspond to the single Dyck word XY,
equivalent to the bracket $\{34\}$,
with the gluon label 5 inserted before, after and in the middle of the word, respectively.
Their color coefficients in the decomposition~\eqref{tree5point}
are given by the following graphs:
\beal \hspace{-10pt}
 & C_{\u{1}\o{2} 5 \u{3}\o{4}} =
      \parbox{80pt}{
      \begin{fmffile}{c12534} \fmfframe(10,3)(0,3){
      \fmfsettings \small
      \begin{fmfgraph*}(60,25)
      \fmfbottom{q2,,,q1}
      \fmftop{l5,,,,q4}
      \fmf{phantom,tension=100}{l5,g5,q3,d3,q4}
      \fmf{phantom,tension=100}{q2,b5,b3,b4,q1}
      \fmf{plain_arrow,tension=0}{q2,b5,b4,q1}
      \fmf{plain_arrow,tension=0}{q4,d3,q3}
      \fmf{curly,tension=0}{b5,g5}
      \fmf{curly,tension=0}{b4,d3}
      \fmfv{label=$\u{1}$,label.angle=0,label.dist=2.5pt}{q1}
      \fmfv{label=$\o{2}$,label.angle=180,label.dist=2.5pt}{q2}
      \fmfv{label=$\u{3}$,label.angle=180,label.dist=2.5pt}{q3}
      \fmfv{label=$\o{4}$,label.angle=0,label.dist=2.5pt}{q4}
      \fmfv{label=$5$,label.angle=90,label.dist=2.5pt}{g5}
      \end{fmfgraph*} }
      \end{fmffile}
      } \,, \hspace{57pt}
   C_{\u{1}\o{2}\u{3}\o{4} 5} =
      \parbox{80pt}{
      \begin{fmffile}{c12345} \fmfframe(10,3)(0,3){
      \fmfsettings \small
      \begin{fmfgraph*}(60,25)
      \fmfbottom{q2,,,q1}
      \fmftop{q3,,,,r5}
      \fmf{phantom,tension=100}{q3,d3,q4,g5,r5}
      \fmf{phantom,tension=100}{q2,b3,b4,b5,q1}
      \fmf{plain_arrow,tension=0}{q2,b3,b5,q1}
      \fmf{plain_arrow,tension=0}{q4,d3,q3}
      \fmf{curly,tension=0}{b3,d3}
      \fmf{curly,tension=0}{b5,g5}
      \fmfv{label=$\u{1}$,label.angle=0,label.dist=2.5pt}{q1}
      \fmfv{label=$\o{2}$,label.angle=180,label.dist=2.5pt}{q2}
      \fmfv{label=$\u{3}$,label.angle=180,label.dist=2.5pt}{q3}
      \fmfv{label=$\o{4}$,label.angle=0,label.dist=2.5pt}{q4}
      \fmfv{label=$5$,label.angle=90,label.dist=2.5pt}{g5}
      \end{fmfgraph*} }
      \end{fmffile}
      } \,, \hspace{-10pt} \\ \hspace{-20pt}
 & C_{\u{1}\o{2}\u{3} 5 \o{4}} =
      \parbox{80pt}{
      \begin{fmffile}{c1c12354} \fmfframe(10,3)(0,3){
      \fmfsettings \small
      \begin{fmfgraph*}(60,50)
      \fmfbottom{q2,,,q1}
      \fmftop{l5,,,r5}
      \fmfleft{q3}
      \fmfright{q4}
      \fmf{phantom,tension=100}{l5,u3,g5,r5}
      \fmf{phantom,tension=100}{q3,d3,d4,q4}
      \fmf{phantom,tension=100}{q2,b3,b4,q1}
      \fmf{plain_arrow,tension=0}{q2,b3,b4,q1}
      \fmf{plain_arrow,tension=0}{q4,d3,q3}
      \fmf{plain_arrow,tension=0}{q4,d4}
      \fmf{curly,tension=0}{b3,d3}
      \fmf{curly,tension=0,rubout=3}{b4,g5}
      \fmfv{label=$\u{1}$,label.angle=0,label.dist=2.5pt}{q1}
      \fmfv{label=$\o{2}$,label.angle=180,label.dist=2.5pt}{q2}
      \fmfv{label=$\u{3}$,label.angle=180,label.dist=2.5pt}{q3}
      \fmfv{label=$\o{4}$,label.angle=0,label.dist=2.5pt}{q4}
      \fmfv{label=$5$,label.angle=90,label.dist=2.5pt}{g5}
      \end{fmfgraph*} }
      \end{fmffile}
      }
      +
      \parbox{80pt}{
      \begin{fmffile}{c4c12354} \fmfframe(10,3)(0,3){
      \fmfsettings \small
      \begin{fmfgraph*}(60,50)
      \fmfbottom{q2,,,q1}
      \fmftop{l5,,,r5}
      \fmfleft{q3}
      \fmfright{q4}
      \fmf{phantom,tension=100}{l5,u3,g5,r5}
      \fmf{phantom,tension=100}{q3,d3,d4,q4}
      \fmf{phantom,tension=100}{q2,b3,b4,q1}
      \fmf{plain_arrow,tension=0}{q2,b3,q1}
      \fmf{plain_arrow,tension=0}{q4,d4,d3,q3}
      \fmf{curly,tension=0}{b3,d3}
      \fmf{curly,tension=0,rubout=3}{d4,g5}
      \fmfv{label=$\u{1}$,label.angle=0,label.dist=2.5pt}{q1}
      \fmfv{label=$\o{2}$,label.angle=180,label.dist=2.5pt}{q2}
      \fmfv{label=$\u{3}$,label.angle=180,label.dist=2.5pt}{q3}
      \fmfv{label=$\o{4}$,label.angle=0,label.dist=2.5pt}{q4}
      \fmfv{label=$5$,label.angle=90,label.dist=2.5pt}{g5}
      \end{fmfgraph*} }
      \end{fmffile}
      }
      =
      \parbox{80pt}{
      \begin{fmffile}{c2c12354} \fmfframe(10,3)(0,3){
      \fmfsettings \small
      \begin{fmfgraph*}(60,50)
      \fmfbottom{q2,,,q1}
      \fmftop{l5,,,r5}
      \fmfleft{q3}
      \fmfright{q4}
      \fmf{phantom,tension=100}{l5,g5,u4,r5}
      \fmf{phantom,tension=100}{q3,d3,d4,q4}
      \fmf{phantom,tension=100}{q2,b3,b4,q1}
      \fmf{plain_arrow,tension=0}{q2,b3,b4,q1}
      \fmf{plain_arrow,tension=0}{q4,d4,q3}
      \fmf{plain_arrow,tension=0}{d3,q3}
      \fmf{curly,tension=0}{b4,d4}
      \fmf{curly,tension=0,rubout=3}{b3,g5}
      \fmfv{label=$\u{1}$,label.angle=0,label.dist=2.5pt}{q1}
      \fmfv{label=$\o{2}$,label.angle=180,label.dist=2.5pt}{q2}
      \fmfv{label=$\u{3}$,label.angle=180,label.dist=2.5pt}{q3}
      \fmfv{label=$\o{4}$,label.angle=0,label.dist=2.5pt}{q4}
      \fmfv{label=$5$,label.angle=90,label.dist=2.5pt}{g5}
      \end{fmfgraph*} }
      \end{fmffile}
      }
      +
      \parbox{80pt}{
      \begin{fmffile}{c3c12354} \fmfframe(10,3)(0,3){
      \fmfsettings \small
      \begin{fmfgraph*}(60,50)
      \fmfbottom{q2,,,q1}
      \fmftop{l5,,,r5}
      \fmfleft{q3}
      \fmfright{q4}
      \fmf{phantom,tension=100}{l5,g5,u4,r5}
      \fmf{phantom,tension=100}{q3,d3,d4,q4}
      \fmf{phantom,tension=100}{q2,b3,b4,q1}
      \fmf{plain_arrow,tension=0}{q2,b4,q1}
      \fmf{plain_arrow,tension=0}{q4,d4,d3,q3}
      \fmf{curly,tension=0}{b4,d4}
      \fmf{curly,tension=0,rubout=3}{d3,g5}
      \fmfv{label=$\u{1}$,label.angle=0,label.dist=2.5pt}{q1}
      \fmfv{label=$\o{2}$,label.angle=180,label.dist=2.5pt}{q2}
      \fmfv{label=$\u{3}$,label.angle=180,label.dist=2.5pt}{q3}
      \fmfv{label=$\o{4}$,label.angle=0,label.dist=2.5pt}{q4}
      \fmfv{label=$5$,label.angle=90,label.dist=2.5pt}{g5}
      \end{fmfgraph*} }
      \end{fmffile}
      } \,, \hspace{-20pt}
\label{cprimitives5point}
\eeal
where $C_{\u{1}\o{2}\u{3} 5 \o{4}}$
is drawn both as $(-c_1+c_4)$ and $(-c_2+c_3)$
to emphasize that the nonplanar diagrams cannot be removed
by commutation relations.
The pattern to take note of is that the non-planarity occurs in the ordering $\{3\:5\:4\}$ with the gluon sandwiched between the quark brackets,
which is reminiscent of the nested quark-antiquark pairs in \sec{sec:multiquark}.

%%%%%%%%%%%%%%%%%%%%%%%%%%%%%%%%%%%%%%%%%%%%%%%%%%%%
\subsection{New color decomposition}
\label{sec:colorfactors}
%%%%%%%%%%%%%%%%%%%%%%%%%%%%%%%%%%%%%%%%%%%%%%%%%%%%

In this section we formulate the new color decomposition for QCD.

We use the Melia basis for primitives with $(n-2k)$ gluons
and $k$ quark-lines~\cite{Melia:2013epa}:
\be
   \big\{ A(\u{1},\o{2},\sigma) ~\big|~
          \sigma \in \text{Dyck}_{k-1}
          \times \{\text{gluon insertions}\}_{n-2k} \big\} \,.
\label{KKbasis}
\ee
In the construction of this basis there are $(2k-2)!/(k!(k-1)!)$ Dyck words prior to assigning the particle labels inside the brackets.
The quark labels can be assigned to $(k-1)$ slots
in $(k-1)!$ inequivalent ways.
The antiquark labels have unique slot assignments after this,
since all quark lines have different flavors.
Then the $(n-2k)$ gluons are assigned to any place except
between $\u{1}$ and $\o{2}$, which must stay adjacent.
With each gluon inserted, the number of available slots grows,
starting from $(2k-1)$ up to $(n-2)$.
Therefore, the size of this color-algebra basis is
\be
   \varkappa(n,k) =
   \underbrace{ \overbrace{\frac{(2k-2)!}{k!(k-1)!}}^\text{empty brackets}
                \times (k-1)!}_\text{dressed quark brackets}
      \times \underbrace{(2k-1)(2k)\dots(n-2)}_{\text{insertions of }
                                                 (n-2k)\text{ gluons}}
      = \frac{(n-2)!}{k!} \,,
\label{KKbasisLength}
\ee
in agreement with the reasoning given in the beginning of \sec{sec:coloralgebra}.
See \tab{tab:IndependentKK}
for the explicit counts of the lower-multiplicity primitive amplitudes.

%%%%%%%%%% TABLE %%%%%%%%%%
\begin{table*}
\centering
\begin{tabular}{|c||c|c|c|c|c|c|}
\hline
$k \setminus n$ & 3 & 4 & 5 & 6 & 7 & 8 \\
\noalign{\hrule height 0.7pt}
0 & 1 & 2 & 6 & 24 & 120 & 720 \\
\hline
1 & 1 & 2 & 6 & 24 & 120 & 720 \\
\hline
2 & - & 1 & 3 & 12 &  60 & 360 \\
\hline
3 & - & - & - &  4 &  20 & 120 \\
\hline
4 & - & - & - &  - &   - &  30 \\
\hline
\end{tabular}
%\\ \vskip 10pt
%$\varkappa(n,k)=\frac{(n-2)!}{k!}$ for $2k \le n$
\caption{\small The number of $n$-point primitive amplitudes
         with $k$ distinguishable quark pairs,
         independent under color-algebra relations, as given by the formula
         $\varkappa(n,k) = (n-2)!/k!$.}
\label{tab:IndependentKK}
\end{table*}
%%%%%%%%%%%%%%%%%%%%%%%%%%%

Now the new color decomposition for QCD is conveniently written as
\be
   {\cal A}^{\rm tree}_{n,k} = \!\!\! 
      \sum_{\sigma\:\!\in\text{\,Melia\;basis}}^{\varkappa(n,k)} \!\!
      C(\u{1},\o{2},\sigma) \, A(\u{1},\o{2},\sigma) \,,
\label{NewColorDecomposition}
\ee
where $A(\u{1},\o{2},\sigma)$ are the usual color-ordered planar primitive amplitudes (defined by the color-ordered Feynman rules in \app{app:FeynmanRules}), and the color coefficients $C(\u{1},\o{2},\sigma)$ are nontrivial objects that are the subject of the remaining discussion in this section. 

\begin{figure}[t]
\centering
\vspace{7pt}
\includegraphics[scale=1.0,trim=0 0 0 0,clip=true]{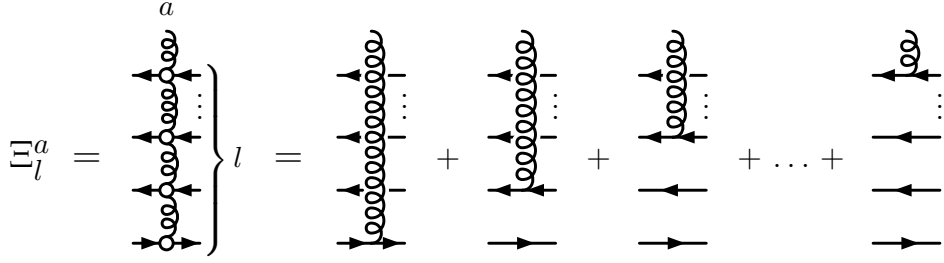}
\caption{\small Diagrammatic form of the operator $\Xi^a_l$.
         It is drawn as a single diagram with hollow quark-gluon vertices, this represents summation
         over the possible locations where the gluon line can attach.}
\label{fig:Xi}
\end{figure}

Using the suggestive bracket notation,
we can obtain the color coefficients using the following replacement rules
for the quark, antiquark and gluon labels:
\be
   C(\u{1},\o{2},\sigma) = (-1)^{k-1}\, \{2|\sigma|1\} {\Bigg| \scriptsize
      \begin{array}{l} \\
         \u{q}~\,\rightarrow\,\{q| \, T^b\!\otimes \Xi_{l-1}^b \\
         \o{q}~\,\rightarrow~|q \} \\
         g~\,\rightarrow~\Xi_{l}^{a_g}
      \end{array} } \,,
\label{KKcolor}
\ee
where the integer $l$ is the level of bracket ``nestedness''
for a given particle in the word $\{2|\sigma|1\}$.
In other words, $l$ is the number of opening brackets minus the number of closing brackets to the left of the particle.
The bra $\{q|$ and the ket $|q\}$ now represent the fundamental and anti-fundamental color indices of a quark and an antiquark,
or can equivalently be understood as their the color wavefunctions.
The object $\Xi_{l}^a$ in \eqn{KKcolor} is an operator obtained by tensoring $l$ copies of the Lie algebra,
\be
   \Xi_{l}^a \, = \, \sum_{s=1}^{l}\,
      \underbrace{\,1 \otimes \cdots \otimes 1 \otimes
         \overbrace{T^a \otimes 1 \otimes \cdots \otimes 1 \otimes \o{1}\,}^{s}
                 }_{l}\,.
\label{Xidef}
\ee
The sum effectively runs over each slot $s$ in the identity tensor product 
and inserts a generator with adjoint index $a$ and in the appropriate representation, see \fig{fig:Xi}.
By convention, each copy of the Lie algebra representation corresponds
to a particular nestedness level, starting from level $l$ (the leftmost copy) and down to level one (the rightmost copy).
Note that the operators $\Xi_{l}^a$ form a representation of the Lie algebra,
\be
   \big[\Xi_{l}^a,\,\Xi_{l}^b \big]=\tf^{abc}\,\Xi_{l}^c\,.
\ee
This makes it natural to extend \eqn{Xidef} to the pure-gluon case
by defining $\Xi_{0}^a=T^a_\text{adj}$.

In the multi-sandwich formula~\eqref{KKcolor}
the quark and antiquark wavefunctions act only on the Lie algebra copy
at their corresponding nestedness level $l$.
For example, $\{2|$ and $|1\}$ act on the level-one copy of the group representation, which is complex conjugated with respect to the rest, as is seen in~\fig{fig:Xi}.
This is indicated by the bar over the rightmost unit operator in \eqn{Xidef}
and the fact that in \eqn{KKcolor} we use the bra $\{2|$ and the ket $|1\}$
for the color wavefunctions of the antiquark $\o{2}$ and quark $\u{1}$,
which is contrary to the convention for all other particles.
This is due to the special role of the fermion line $\u{1} \leftarrow \o{2}$.
This notational subtlety could in principle be avoided by complex conjugating
that line into $\o{1} \rightarrow \u{2}$,
\ie by interchanging the roles of the quark and the antiquark.\footnote{This
would mean the Melia basis $A(\o{1},\u{2},\sigma)$,
or, after relabeling, $A(\u{1},\sigma,\o{n})$.} More generally,
any group representation copy can be complex conjugated,
as it only changes the interpretation of which particles are quarks and antiquarks. (In fact, any representation can be allowed
for each Lie algebra copy.)

%%%%%%%%% FIGURE %%%%%%%%%%%%%%%
%%%%%%%%%%%%%%%%%%%%%%%%%%%%%%%%

The above formulas~\eqref{KKcolor} and~\eqref{Xidef}
unambiguously determine the color coefficients
in the color decomposition~\eqref{NewColorDecomposition}.
For $k=1,0$ it is straightforward to see that our decomposition
coincides with the multi-peripheral formulas~\eqref{QuarkLineDecomposition}
and~\eqref{DDM}.
Indeed, if all permuted particles $3,\dots,n$ are gluons,
they are replaced by $\Xi^{a_i}_1 = \o{T}^{a_i}$,
producing precisely~\eqn{QuarkLineDecomposition}.
In case particles~1 and~2 are gluons as well, the nestedness level is zero,
and the adjoint-representation operators $\Xi^{a_i}_0 = T^{a_i}_\text{adj}$
yield the DDM decomposition (\ref{DDM}).

For a more detailed example of using \eqn{KKcolor},
let us consider the six-point amplitude
discussed in \sec{sec:multiquark}.
One of the color coefficients in its decomposition is
\beal
   C_{\u{1}\o{2}\u{3}\o{4}\u{5}\o{6}} & =
      \{2| \{3| T^a\!\otimes \Xi_1^a |4\}
           \{5| T^b\!\otimes \Xi_1^b |6\} |1\} =
      \{2| \{3| T^{a}\!\otimes \o{T}^{a} |4\}
           \{5| T^{b}\!\otimes \o{T}^{b} |6\} |1\} \\ & =
      \{2| \o{T}^{a} \o{T}^{b} |1\} \{3| T^{a} |4\}  \{5| T^{b} |6\} = 
      (T^{b} T^{a})_{i_1 \o{\imath}_2} T^{a}_{i_3 \o{\imath}_4}
                                       T^{b}_{i_5 \o{\imath}_6} \,,
\label{cprimitives6point1}  
\eeal
where we in the last step have translated it to a more conventional notation.
A more interesting example is the color coefficient
\begin{align}
C_{\u{1}\o{2}\u{3}\u{5}\o{6}\o{4}} & =
      \{2| \{3| T^a\!\otimes \Xi_1^a
           \{5| T^b\!\otimes \Xi_2^b |6\} |4\} |1\} \nn \\
  & = \{2| \{3| T^{a}\!\otimes \o{T}^{a}
           \{5| T^{b}\!\otimes 1 \otimes \o{T}^{b} |6\} |4\} |1\} \,+\,
      \{2| \{3| T^{a}\!\!\otimes \o{T}^{a}
           \{5| T^{b}\!\otimes T^{b}\!\otimes \o{1}\,|6\} |4\} |1\} \nn \\
  & = \{2| \o{T}^{a} \o{T}^{b} |1\} \{3| T^{a} |4\}  \{5| T^{b} |6\} \,+\,
      \{2| \o{T}^a |1\} \{3| T^{a} T^{b} |4\}  \{5| T^{b} |6\} \\
  & = (T^{b}T^{a})_{i_1 \o{\imath}_2} T^{a}_{i_3 \o{\imath}_4}
                                      T^{b}_{i_5 \o{\imath}_6} \,-\,
      T^{a}_{i_1 \o{\imath}_2} (T^{a}T^{b})_{i_3 \o{\imath}_4}
                                      T^{b}_{i_5 \o{\imath}_6} \,. \nn
\label{cprimitives6point2}
\end{align}
The other two color coefficients for the six-point amplitude are, with less details, given by
\beal
   C_{\u{1}\o{2}\u{5}\o{6}\u{3}\o{4}} =
      \{2| \{5| T^a\!\otimes \Xi^a_1 |6\}
           \{3| T^b\!\otimes \Xi_1^b |4\} |1\} & =
      \{2| \o{T}^{a} \o{T}^{b} |1\} \{3| T^{b} |4\} \{5| T^{a} |6\} \,,  \\
   C_{\u{1}\o{2}\u{5}\u{3}\o{4}\o{6}} =
      \{2| \{5| T^a\!\otimes \Xi^a_1
           \{3| T^b\!\otimes  \Xi^b_2 |4\} |6\} |1\} & =
      \{2| \o{T}^{a} \o{T}^{b} |1\} \{3| T^{b} |4\} \{5| T^{a} |6\} \\ 
   & \quad +\,\{2| \o{T}^a |1\} \{3| T^{b} |4\} \{5| T^{a} T^{b} |6\} \,.
\label{cprimitives6point34}
\eeal
The four above color coefficients are indeed the ones given diagrammatically
in \eqn{cprimitives6point}.

For the five-point amplitude discussed in \sec{sec:5pointcolor},
our decomposition results in the following color coefficients:
\begin{align}
   C_{\u{1}\o{2} 5 \u{3}\o{4}} & =
      -\, \{2| \Xi^{a_5}_1 \{3| T^b \!\otimes \Xi^b_1 |4\} |1\} =
      -\, \{2| \o{T}^{a_5} \{3| T^{b} \!\otimes \o{T}^{b} |4\} |1\} =
      -\, \{2| \o{T}^{a_5}\o{T}^{b}|1\} \{3| T^{b} |4\} \,, \nn \\
   C_{\u{1}\o{2}\u{3}\o{4} 5} & =
      -\, \{2| \{3| T^b\!\otimes \Xi^b_1  |4\} \Xi^{a_5}_1 |1\} =
      -\, \{2| \{3| T^{b}\!\otimes \o{T}^{b} |4\} \o{T}^{a_5} |1\} =
      -\, \{2| \o{T}^{b}\o{T}^{a_5} |1\} \{3| T^{b} |4\} \,, \nn \\
    C_{\u{1}\o{2}\u{3} 5 \o{4}} & =
      -\, \{2| \{3| (T^b\!\otimes \Xi^b_1) \, \Xi^{a_5}_2 |4\} |1\} \\ & =
      -\, \{2| \{3| (T^b\!\otimes \o{T}^{b}) (1 \otimes \o{T}^{a_5}) |4\} |1\}
    \,-\, \{2| \{3| (T^b\!\otimes \o{T}^{b})
                    (T^{a_5}\!\otimes \o{1}\,) |4\} |1\} \nn \\ & =
      -\, \{2| \o{T}^{b} \o{T}^{a_5} |1\} \{3| T^b |4\}
    \,-\, \{2| \o{T}^{b} |1\} \{3|T^b T^{a_5} |4\} \,, \nn
\label{cprimitives5point}
\end{align}
which coincide with the color diagrams in \eqn{cprimitives5point}.

While we do not provide a proof for the decomposition~\eqref{NewColorDecomposition},
we have explicitly checked its validity for all quark-gluon configurations up to eight points, as well as the nine-point amplitude with four quark lines and one gluon. To be more precise, the check was done as follows.
We expand the Melia basis of color-ordered primitives in kinematic cubic diagrams,
\be
   A(\u{1},\o{2},\sigma) = \!\!\!
      \sum_{\sigma\text{-color-ordered cubic graphs }\Gamma_i} \!\!\!
      \pm\,\frac{n_i}{D_i} \,,
\label{CubicPrimitive}
\ee
and solve for $n_i/D_i$ in terms of the primitives $A(\u{1},\o{2},\sigma)$. When this solution is plugged back into the color dressed cubic graph expansion~\eqref{BCJformYM}, the color coefficients of the undetermined $n_i$
can be shown to vanish under the color algebra~\eqref{coloralgebra}.
What is left is then an expansion of the amplitude
in terms of the primitives and some color factors $c_i$.
The color coefficients $C(\u{1},\o{2},\sigma)$ of the primitives
can then be read off from this expression.
They are indeed given exactly by \eqn{KKcolor}.

%%%%%%%%%%%%%%%%%%%%%%%%%%%%%%%%%%%%%%%%%%%%%%%%%%%%
\subsection{Higher-point example and color-coefficient diagram}
\label{sec:14point}
%%%%%%%%%%%%%%%%%%%%%%%%%%%%%%%%%%%%%%%%%%%%%%%%%%%%

Here we illustrate how the general formula~\eqref{KKcolor}
is applied to a high-multiplicity example. Consider the color factor of the following
14-point primitive amplitude with six quark pairs and two gluons $(n=14, k=6)$:
\be
   A(\u{1},\o{2},13,\u{3},\u{5},\o{6},\o{4},
     \u{7},\u{9},14,\u{11},\o{12},\o{10},\o{8})\,.
\label{primitive14point}
\ee
We can use the cyclic property of planar amplitudes to move $\u{1}$ to the end and then, to make the nestedness of the primitive more apparent, replace the bar notation of the legs with brackets:
\be
\{2\:13 \{ 3 \{5\:6\} 4 \} \{ 7 \{ 9\:14 \{11\:12\} 10 \} 8 \} 1\}\,.
\label{word14}
\ee
As before, the bar-bracket correspondence for the fermion line
$\u{1}\leftarrow\o{2}$ is opposite to all other legs due to the different convention for Lie algebra representation on that line.

Using \eqn{KKcolor} it is straightforward to obtain the expression for the color coefficient of this primitive
\begin{align}
\label{color14point}
   C_{\u{1},\o{2},13,\u{3},\u{5},\o{6},\o{4},
      \u{7},\u{9},14,\u{11},\o{12},\o{10},\o{8}} =
   -\,\{2| \Xi^{a_{13}}_1 & \{3|T^b\!\otimes \Xi^b_1
           \{5| T^c\!\otimes \Xi^c_2 |6\} |4\} \\ \times &
      \{7|T^d\!\otimes \Xi^d_1
           \{9| (T^e\!\otimes \Xi^e_2) \Xi^{a_{14}}_3
                \{11| T^f\!\!\otimes \Xi^f_3 |12\} |10\} |8\}
   |1\} \,. \nn
\end{align}

While this formula is compact, it becomes a rather formidable expression
if written in terms of standard color factors. In order to understand it better, let us sketch out the various standard color structures contained in it. 
For example, the first $\Xi^{a_{13}}_1$ in \eqn{color14point} will give rise to the structure
\be
   \{2| \Xi^{a_{13}}_1 \dots |1\}
      = \big(\!\;\o{T}^{a_{13}} \dots \big)_{\bar \imath_2 i_1} \,,
\ee
whereas the second appearance of $\Xi$ gives
\be
   \{2| \dots \{3|T^{a}\!\otimes \Xi^{a}_1 \dots |4\} \dots |1\}
      = \big( T^{a} \dots \big)_{i_3 \bar \imath_4}
        \big( \dots \o{T}^{a} \dots \big)_{\bar \imath_2 i_1} \,.
\ee
For $l\ge2$ the operator $\Xi^a_l$ consists of a sum of tensor products,
thus nested curly brackets imply summation over different possibilities
of inserting the generator $T^a$.
For example, the third $\Xi$ in \eqn{color14point} gives rise to two structures
\beal
   \{2| \dots \{3 \dots \{5| T^{b}_3\!\otimes \Xi^{b}_2 |6\}
              |4\} \dots |1\} = \,
    & \big( T^b \big)_{i_5 \bar \imath_6}
      \big( \dots T^b \big)_{i_3 \bar \imath_4}
      \big( \dots\,\dots \big)_{\bar \imath_2 i_1} \\ + \,
    & \big( T^b \big)_{i_5 \bar \imath_6}
      \big( \dots \big)_{i_3 \bar \imath_4}
      \big( \dots \o{T}^b \dots \big)_{\bar \imath_2 i_1} \,,
\eeal
while the sixth occurrence of $\Xi$ gives three contributions
\beal
   \{2| \dots \{7 \dots \{9|\dots \Xi^{a_{14}}_3 \dots |10\} |8\} |1\} = \,
    & \big( \dots T^{a_{14}} \dots \big)_{i_9 \bar \imath_{10}}
      \big( \dots \big)_{i_7 \bar \imath_8}
      \big( \dots \big)_{\bar \imath_2 i_1} \\ + \,
    & \big( \dots \, \dots \big)_{i_9 \bar \imath_{10}}
      \big( \dots T^{a_{14}} \big)_{i_7 \bar \imath_8}
      \big( \dots \big)_{\bar \imath_2 i_1} \\ + \,
    & \big( \dots \, \dots \big)_{i_9 \bar \imath_{10}}
      \big( \dots \big)_{i_7 \bar \imath_8}
      \big( \dots \o{T}^{a_{14}} \big)_{\bar \imath_2 i_1} \,.
\eeal
Indeed, each $\Xi^a_l$ gives rise to exactly $l$ structures, which then are multiplied together. This implies that the number of standard color factors hiding in \eqn{color14point} can be counted by multiplying the subscripts of the $\Xi$'s in this expression, giving
\be
   1 \times 1 \times 2 \times 1 \times 2 \times 3 \times 3 = 36~\text{terms}.
\ee

%%%%%%%%% FIGURE %%%%%%%%%%%%%%%
\begin{figure}[t]
\centering
\vspace{7pt}
\includegraphics[scale=1.0,trim=0 0 0 0,clip=true]{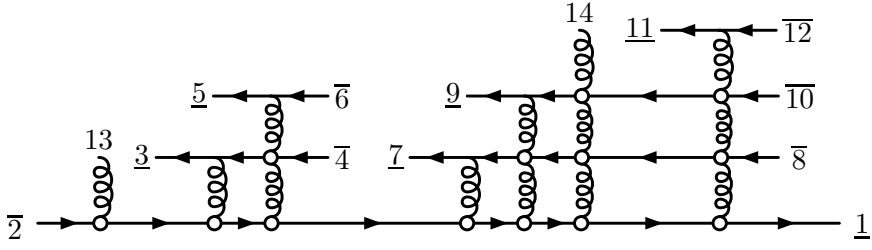}
\caption{\small Diagrammatic representation for the color coefficient
         of the planar amplitude
         $A(\u{1},\o{2},13,\u{3},\u{5},\o{6},\o{4},
            \u{7},\u{9},14,\u{11},\o{12},\o{10},\o{8})$,
         obtained by using the notation of \fig{fig:Xi}.
         Note that the diagram has the same structure as the word
         $\{2\:13 \{ 3 \{5\:6\} 4 \} \{ 7 \{ 9\:14 \{11\:12\} 10 \} 8 \} 1\}$.}
\label{fig:colorfactor}
\end{figure}
%%%%%%%%%%%%%%%%%%%%%%%%%%%%%%%%

Finally, we note that probably the best way to understand the color coefficient~\eqref{color14point} is to draw a diagram for it. Indeed, \fig{fig:colorfactor} contains the same information as the formula~\eqref{color14point}. In particular, compare this diagram with the word given in \eqn{word14}. The diagram \fig{fig:colorfactor} is similar to a usual color factor diagram that describes the contractions of $\tf^{abc}$'s and $T^{a}$'s; however, we use hollow quark-gluon vertices to represent the operation of summing over the possible locations where a gluon can attach (see \fig{fig:Xi}).
In this sum, if the gluon does not attach at a given vertex, the vertex should be considered to be the identity operator. In this sense, the hollow vertex can either represent a gluon attachment, or a nonplanar crossing of the lines. Note that the diagram in \fig{fig:colorfactor}, and similar ones built up by the diagrammatic $\Xi$'s shown in \fig{fig:Xi}, are the natural generalizations of the multi-peripheral diagrams that appear in the DDM decomposition.

%%%%%%%%%%%%%%%%%%%%%%%%%%%%%%%%%%%%%%%%%%%%%%%%%%%%
\section{Kinematic-algebra basis for quark-gluon amplitudes}
\label{sec:kinematicalgebra}
%%%%%%%%%%%%%%%%%%%%%%%%%%%%%%%%%%%%%%%%%%%%%%%%%%%%

In this section we turn our attention to the kinematic structure
of general tree amplitudes in QCD.
We find that they obey the color-kinematics duality (\ref{duality}),
and this constrains the basis of primitives further. The number of independent primitives is reduced down
to $(n-3)!(2k-2)/k!$ for $k\ge2$, and for $k=1,0$ it is given by the familiar $(n-3)!$.

The color-kinematics duality is well-established for gluons in any spacetime dimension~\cite{Bern:2008qj,Bern:2010yg,Bern:2013yya}, and similarly for massless particles in supersymmetric YM multiplets that comprise gluons~\cite{Bern:2010ue,Carrasco:2011mn,Bern:2011rj,Bern:2012uf,Carrasco:2012ca,Bjerrum-Bohr:2013iza,Bern:2013uka,Bern:2014sna} or only matter~\cite{Nohle:2013bfa,Chiodaroli:2013upa,Johansson:2014zca}.
The BCJ amplitude relations, which are a consequence of the duality~\cite{Bern:2008qj}
but can be proven by other means~\cite{BjerrumBohr:2009rd,Stieberger:2009hq,Feng:2010my},
have in a few cases been shown to generalize to massive amplitudes. In particular, amplitudes with a massive scalar pair~\cite{Naculich:2014naa}
and with three massive particles of different spins~\cite{Naculich:2015zha} have been considered in the literature.

In this section, we show that the color-kinematics duality (\ref{duality}) holds
for QCD tree-level $n$-point amplitudes with $k$ quark-antiquark pairs having distinct flavors and masses. We prove this for $k\le 4$, $n\le 8$ by explicit calculations. 
Assuming the duality holds in general, we derive the resulting tree-amplitude relations.
We expect that these relations not only hold for QCD, but can be applied to generic YM amplitudes with massive flavored matter particles (fermions/scalars), including general gauge-group representation, general spacetime dimension, with and without supersymmetry. In particular, the amplitude relations can be used as a gauge-invariant test to check if a given theory obeys the color-kinematics duality.

We first give a few concrete examples and then state the general results.

%%%%%%%%%%%%%%%%%%%%%%%%%%%%%%%%%%%%%%%%%%%%%%%%%%%%
\subsection{Four-point example: $n=4$, $k=1$}
\label{sec:4point}
%%%%%%%%%%%%%%%%%%%%%%%%%%%%%%%%%%%%%%%%%%%%%%%%%%%%

One of the simplest massive quark amplitudes involves
one massive quark-antiquark pair and two gluons.
It has three Feynman diagrams:
\begin{subequations} \begin{align}
      \parbox{80pt}{
      \begin{fmffile}{QQgg1} \fmfframe(10,10)(0,10){
      \fmfsettings
      \begin{fmfgraph*}(50,25)
            \fmflabel{$\u{1}, i$}{q1}
            \fmflabel{$\o{2}, \bar \jmath$}{q2}
            \fmflabel{$3, a\!\!\!$}{g3}
            \fmflabel{$\!\!\!\!4, b$}{g4}
            \fmftop{,g3,,g4,}
            \fmfbottom{q1,q2}
            \fmf{plain_arrow}{q2,v4,v3,q1}
            \fmf{curly,tension=0}{g3,v3}
            \fmf{curly,tension=0}{g4,v4}
      \end{fmfgraph*} }
      \end{fmffile}
      }
      & = -\frac{i}{2} \,
          \frac{ T_{i \bar k}^a T_{k \bar \jmath}^b }
               { s_{13}\!-\!m^2 } \,
        ( \bar{u}_1 {\not}\varepsilon_3 ({\not}k_{1,3}\!+\!m)
                    {\not}\varepsilon_4 v_2 )
        =  \frac{c_1 n_1}{D_1} \,,
\label{QQgg1} \\
      \parbox{80pt}{
      \begin{fmffile}{QQgg2} \fmfframe(10,10)(0,10){
      \fmfsettings
      \begin{fmfgraph*}(50,25)
            \fmflabel{$\u{1}, i$}{q1}
            \fmflabel{$\o{2}, \bar \jmath$}{q2}
            \fmflabel{$4, b\!\!\!$}{g4}
            \fmflabel{$\!\!\!\!3, a$}{g3}
            \fmftop{,g4,,g3,}
            \fmfbottom{q1,q2}
            \fmf{plain_arrow}{q2,v3,v4,q1}
            \fmf{curly,tension=0}{g3,v3}
            \fmf{curly,tension=0}{g4,v4}
      \end{fmfgraph*} }
      \end{fmffile}
      }
      & = -\frac{i}{2} \,
          \frac{ T_{i \bar k}^b T_{k \bar \jmath}^a }
               { s_{14}\!-\!m^2 } \,
        ( \bar{u}_1 {\not}\varepsilon_4 ({\not}k_{1,4}\!+\!m)
                    {\not}\varepsilon_3 v_2 )
        =  \frac{c_2 n_2}{D_2} \,,
\label{QQgg2} \\
      \parbox{80pt}{
      \begin{fmffile}{QQgg3} \fmfframe(10,0)(0,0){
      \fmfsettings
      \begin{fmfgraph*}(50,40)
            \fmflabel{$\u{1}, i$}{q1}
            \fmflabel{$\o{2}, \bar \jmath$}{q2}
            \fmflabel{$3, a\!\!\!$}{g3}
            \fmflabel{$\!\!\!\!4, b$}{g4}
            \fmftop{,g3,,,,g4,}
            \fmfbottom{q1,q2}
            \fmf{plain_arrow}{q2,v12,q1}
            \fmf{curly,tension=0.015}{v12,v34}
            \fmf{curly,tension=0.01}{g3,v34}
            \fmf{curly,tension=0.01}{g4,v34}
      \end{fmfgraph*} }
      \end{fmffile}
      } & = \!
      \begin{aligned} \\
          \frac{i}{2}
          \frac{ \tilde{f}^{abc} T_{i \bar \jmath}^c }
               { s_{12} } \,
          \Big( 2 (k_{4}\!\cdot\!\varepsilon_3)
                  ( \bar{u}_1 {\not}\varepsilon_4 v_2 )
              - 2 (k_{3}\!\cdot\!\varepsilon_4)
                  ( \bar{u}_1 {\not}\varepsilon_3 v_2 ) & \\
              +\, (\varepsilon_3\!\cdot\!\varepsilon_4)
                  ( \bar{u}_1 ({\not}k_3-{\not}k_4) v_2 ) & \Big)
        = \frac{c_3 n_3}{D_3} \,,
      \end{aligned}\!\!\!\!\!\!\!\!\!\!
\label{QQgg3}
\end{align} \label{QQgg}%
\end{subequations}
where the spins are left unspecified.
These diagrams correspond precisely
to the ones in the fundamental commutation relation~\eqref{commutation}, which in terms of color factors read
\be
   c_1 - c_2 = c_3 \,.
\ee
By the color-kinematics duality, the kinematic relation should then read
\be
   n_1 - n_2 = n_3 \,.
\label{duality4point}
\ee
Indeed, the kinematic numerators
of the above Feynman diagrams,
\beal
 & n_1 = -\frac{i}{2} \,
          \bar{u}_1 {\not}\varepsilon_3 ({\not}k_{1,3}\!+\!m)
                    {\not}\varepsilon_4 v_2 \,, \qquad \qquad \quad\:\,
   n_2 = -\frac{i}{2} \,
          \bar{u}_1 {\not}\varepsilon_4 ({\not}k_{1,4}\!+\!m)
                    {\not}\varepsilon_3 v_2 \,, \\
 & n_3 =  \frac{i}{2}
          \Big( 2 (k_{4}\!\cdot\!\varepsilon_3)
                  ( \bar{u}_1 {\not}\varepsilon_4 v_2 )
              - 2 (k_{3}\!\cdot\!\varepsilon_4)
                  ( \bar{u}_1 {\not}\varepsilon_3 v_2 )
     +\, (\varepsilon_3\!\cdot\!\varepsilon_4)
                  ( \bar{u}_1 ({\not}k_3-{\not}k_4) v_2 ) \Big) \,,
\label{numerators4point}
\eeal
do satisfy \eqn{duality4point}. To check this in detail one has to repeatedly use the Clifford algebra,
\be
    \gamma^\mu \gamma^\nu + \gamma^\nu \gamma^\mu = 2 \eta^{\mu \nu} \,,
\ee
the Dirac equations and gluon transversality conditions,
\be
   \bar{u}_1 ({\not}k_1\!-\!m) = 0 \,, \qquad 
   ({\not}k_2\!+\!m) v_2 = 0 \,, \qquad k_i \cdot \varepsilon_i=0\,,
\ee
as well as the mass-shell conditions for the quarks $k_1^2=k_2^2=m^2$ and the gluons $k_3^2=k_4^2=0$. 
After several algebraic steps, one arrives at
\be
   n_1 - n_2 - n_3 \,\propto ~
         \bar{u}_1 {\not}k_1 {\not}\varepsilon_3 {\not}\varepsilon_4 v_2
      +  \bar{u}_1 {\not}\varepsilon_3 {\not}\varepsilon_4 {\not}k_2 v_2
      - (\varepsilon_3\!\cdot\!\varepsilon_4)
        (\bar{u}_1 ({\not}k_1+{\not}k_2) v_2) = 0 \,.
\ee

Note that the numerators~\eqref{numerators4point} are gauge-dependent
through the polarization vectors of the gluons,
but the combination $n_1 - n_2 - n_3$ is gauge-invariant (and zero).
For a generic amplitude the numerators of cubic graphs need to absorb the four-gluon interactions, which invariably leads to ambiguities in defining the numerators, and thus the duality is not manifest in general. 
However, in the above amplitude there is no quartic vertex, hence it is not surprising that the duality holds automatically from the Feynman rules. 

Having shown that the color-kinematics duality is present in this QCD amplitude, we proceed using only the formal properties of the numerators. Expressing the color-dressed amplitude through two color-ordered primitives
is straightforward,
\be
   {\cal A}_{4,1}^\text{tree} = \sum_{i=1}^3 \frac{c_i n_i}{D_i}
    = \Bigg\{ c_1 \left( \frac{n_1}{D_1} + \frac{n_3}{D_3} \right)
            + c_2 \left( \frac{n_2}{D_2} - \frac{n_3}{D_3} \right)
      \Bigg\} \\
      \equiv  c_2 A_{\u{1}\o{2} 3 4} + c_1 A_{\u{1}\o{2} 4 3} \,,
\label{tree4point}
\ee
where the color factors are of the multi-peripheral type, in accord with
the decomposition~\eqref{QuarkLineDecomposition}.
Using the duality, $n_3=n_1-n_2 $, we obtain a system of two equations,
\be
   A_{\u{1}\o{2} 3 4} = \left( \frac{1}{D_2} + \frac{1}{D_3} \right) n_2
                      - \frac{n_1}{D_3} \,, \qquad \quad
   A_{\u{1}\o{2} 4 3} = \left( \frac{1}{D_1} + \frac{1}{D_3} \right) n_1
                      - \frac{n_2}{D_3} \,,
\label{primitives4point}
\ee
which, after Gaussian elimination of $n_1$, yields
\be
   A_{\u{1}\o{2} 3 4} = \left( \frac{1}{D_2} + \frac{1}{D_3}
                             - \frac{D_1}{(D_1\!+\!D_3) D_3} \right) n_2
                      - \frac{D_1}{D_1\!+\!D_3} A_{\u{1}\o{2} 4 3} \,.
\ee
The coefficient of $n_2$ can be shown to be proportional
to a vanishing sum of denominators
\be
   D_1 + D_2 + D_3 = (s_{13}-m^2) + (s_{14}-m^2) + s_{12} = 0 \,,
\ee
resulting in the following relation among the primitive amplitudes:
\be
   (s_{14}-m^2) A_{\u{1}\o{2} 3 4} = (s_{13}-m^2) A_{\u{1} \o{2} 4 3} \,.
\label{relation4point}
\ee
This is a straightforward generalization\footnote{\Eqn{relation4point} can be presented~\cite{Naculich:2014naa} as $(k_1\cdot k_4) A_{\u{1}\o{2} 3 4} = (k_1\cdot k_3) A_{\u{1} \o{2} 4 3}$, making it formally identical to the massless BCJ relation, but here we prefer to use momentum invariants with explicit mass dependence, similar to Feynman propagators.} of the corresponding massless BCJ relation~(\ref{fBCJ}).
It allows us to express the full amplitude in terms of a single primitive,
\be
   {\cal A}_{4,1}^\text{tree} =
      \Big( T^{a_3}_{\bar \imath_2 j} T^{a_4}_{\bar \jmath\,i_1}
          + T^{a_4}_{\bar \imath_2 j} T^{a_3}_{\bar \jmath\,i_1}
            \frac{s_{14}-m^2}{s_{13}-m^2}
      \Big) A_{\u{1}\o{2} 3 4} \,.
\label{tree4final}
\ee

%%%%%%%%%%%%%%%%%%%%%%%%%%%%%%%%%%%%%%%%%%%%%%%%%%%%
\subsection{Kinematic algebra for $n=5$, $k=2$}
\label{sec:5point}
%%%%%%%%%%%%%%%%%%%%%%%%%%%%%%%%%%%%%%%%%%%%%%%%%%%%

We proceed by showing that the color-kinematics duality is present in QCD amplitudes with two quark lines of different flavors and masses. Let us return to the $n=5$, $k=2$ amplitude considered in \sec{sec:5pointcolor}.
Since the four-gluon vertex is also absent in this amplitude,
the numerators in~\eqn{numerators5point} are directly given by the Feynman rules, and we expect the duality to hold automatically,
\begin{subequations} \begin{align}
 & c_1 - c_2 = - c_5 ~~~~~ \Leftrightarrow ~~~~~
   n_1 - n_2 = - n_5 \,, \\
 & c_3 - c_4 = c_5 ~~~~~~\:\: \Leftrightarrow ~~~~~
   n_3 - n_4 = n_5 \,.
\end{align} \label{duality5point}%
\end{subequations}
As before, to see that these relations indeed hold true, one has to use the Clifford algebra, the Dirac equation for massive quarks, the transversality condition for gluons, as well as the mass-shell conditions. We will spare the reader the details of this successful check, and instead proceed formally.

Combining the two kinematic identities of \eqn{duality5point} and
the three primitives as defined in~\eqn{tree5symmetric},
we obtain a system of five equations.
Four of these equations can be used to solve for numerators $n_1$, $n_2$, $n_3$ and $n_4$ in terms of variables $n_5$, $A_{\u{1} 5 \o{2}\u{3}\o{4}}$ and $A_{\u{1}\o{2}\u{3} 5 \o{4}}$,
\beal
   n_2 = & \: n_1 - n_5 \,, \qquad \quad
   n_1 = \frac{D_1 D_2}{D_1\!+\!D_2}
         (A_{\u{1} 5 \o{2}\u{3}\o{4}} - \frac{n_5}{D_2}) \,, \\
   n_3 = & \: n_4 - n_5 \,, \qquad \quad
   n_4 = \frac{D_3 D_4}{D_3\!+\!D_4}
         (A_{\u{1}\o{2}\u{3} 5 \o{4}} - \frac{n_5}{D_3}) \,,
\label{solvednumerators5point}
\eeal
the remaining fifth equation yields an expression for $A_{\u{1}\o{2}\u{3}\o{4} 5}$,
\be
   A_{\u{1}\o{2}\u{3}\o{4} 5} =
    - \frac{D_2}{D_1\!+\!D_2} A_{\u{1} 5 \o{2}\u{3}\o{4}}
    - \frac{D_3}{D_3\!+\!D_4} A_{\u{1}\o{2}\u{3} 5 \o{4}}
    + \left( \frac{1}{D_1\!+\!D_2}
           + \frac{1}{D_3\!+\!D_4} + \frac{1}{D_5} \right) n_5 \,.
\label{relation5withn5}         
\ee
The variable $n_5$ cannot be solved for since its prefactor in \eqn{relation5withn5} vanishes,
resulting in the following gauge-invariant amplitude relation:
\be
(s_{35}-m_3^2) A_{\u{1}\o{2}\u{3} 5 \o{4}}
+(s_{12} - s_{34}) A_{\u{1}\o{2}\u{3}\o{4} 5}
    -(s_{25}-m_2^2) A_{\u{1} 5 \o{2}\u{3}\o{4}} = 0 \,.
\label{relation5}
\ee
This expression involves the primitive $A_{\u{1} 5 \o{2}\u{3}\o{4}}$
outside of the Melia basis,
but we can trade it for $A_{\u{1}\o{2} 5 \u{3}\o{4}}$
using the KK relation~\eqref{KK5}.
Via the identity $s_{12} - s_{34}= s_{35} + s_{45} - m_3^2 - m_4^2$,
we obtain another amplitude relation:
\beal
   (s_{25} - m_2^2) A_{\u{1}\o{2} 5 \u{3}\o{4}} &
 + (s_{25} + s_{35} - m_2^2 - m_3^2) A_{\u{1}\o{2}\u{3} 5 \o{4}} \\ &
 + (s_{25} + s_{35} + s_{45} - m_2^2 - m_3^2 - m_4^2)
                    A_{\u{1}\o{2}\u{3}\o{4} 5} = 0 \,.
\label{relation5point}
\eeal
Here, in order to have consistent quark masses, one should set $m_1=m_2$ and $m_3=m_4$. We keep the masses distinct, as well as avoid using further momentum identities, so as to match the pattern of the general $n$-point formula given in the next section. If we do use further momentum identities this expression can be simplified to
\be
  (s_{25}-m_2^2) A_{\u{1}\o{2} 5 \u{3}\o{4}}
  + (s_{14} - s_{23}) A_{\u{1}\o{2}\u{3} 5\o{4}} 
   -(s_{15}-m_1^2) A_{\u{1}\o{2}\u{3}\o{4} 5 }
= 0 \,,
\label{relation5simple}
\ee
which is nothing but a relabeling of \eqn{relation5}: $(\u{1},\o{2},\u{3},\o{4})\rightarrow (\o{4},\u{1},\o{2},\u{3})$. In fact, all BCJ relations for $n=5$, $k=2$ are related by such simple relabelings.

As an alternative test of the color-kinematics duality, one can verify these amplitude relations directly, thus circumventing the numerator check. For example, it is straightforward to check that the relations~\eqref{relation5} and~\eqref{relation5simple} hold in the $m_i\rightarrow0$ limit, in which the amplitudes become\footnote{Since $k>1$ these primitives do not correspond to component amplitudes in any of the pure ${\cal N}=1,2,4$ SYM theories. Thus their BCJ relations in principle need not be the same as those of pure SYM theories.}
\beal
\!\!\!
   A(\u{1}^-\!,5^+\!,\o{2}^+\!,\u{3}^-\!,\o{4}^+) &
      = i \frac{ \braket{13}^2 }
               { \braket{15} \braket{25} \braket{34} } \,, \quad
   A(\u{1}^-\!,\o{2}^+\!,5^+\!,\u{3}^-\!,\o{4}^+)
      =  i \frac{ \braket{13}^2 \braket{23} }
                { \braket{12} \braket{34} \braket{25} \braket{35} } \,, \\
\!\!\!
   A(\u{1}^-\!,\o{2}^+\!,\u{3}^-\!,5^+\!,\o{4}^+) &
      =  i \frac{ \braket{13}^2 }
                { \braket{12} \braket{35} \braket{45} } \,, \quad
   A(\u{1}^-\!,\o{2}^+\!,\u{3}^-\!,\o{4}^+\!,5^+)
      = -i \frac{ \braket{13}^2 \braket{14} }
                { \braket{12} \braket{34} \braket{15} \braket{45} } \,,
\label{primitives5massless}
\eeal
where we have specialized to quarks and a gluon of definite helicity, using the spinor-helicity formalism~\cite{Elvang:2013cua}. Of course, the above amplitude relations hold just as well for any mass and spin/helicity configuration, as follows from the fact that the Feynman diagrams in~\eqn{numerators5point} satisfy the color-kinematics duality.

We note that \eqn{relation5point} can be mapped
to the gluonic BCJ relation~\eqref{fBCJ} by letting $s_{j5} \rightarrow (s_{j5}-m_j^2)=2 k_j \cdot k_5$. However, not all of the gluonic BCJ relations hold true for quark primitives. For example,
consider the following identity valid for gluons~\cite{Bern:2008qj}:
\be
   A_{14235} = -\frac{s_{12} s_{45} A_{12345}
                    +s_{25} (s_{14} + s_{24}) A_{15234} }
                    { s_{35} s_{24} } \,.
\ee
In trying to map the primitives to the quark flavor/mass configurations considered here, the left-hand side involves crossed flavor lines and thus should be set to zero.
The resulting two-term relation between
$A_{\u{1}\o{2}\u{3}\o{4} 5}$ and $A_{\u{1} 5 \o{2}\u{3}\o{4}}$ is not a true identity, not even in the massless limit.
This shows that mapping the gluonic BCJ relations onto the $k\ge2$ quark case
is more subtle than for the corresponding KK relations,
which we argued in \sec{sec:coloralgebra} to hold true regardless of the particle configuration.
Nevertheless, it is not difficult to understand the general structure of the quark-gluon BCJ relations, as we show in \sec{sec:bcj}.

Before proceeding to higher-point amplitudes, we note that one can use \eqn{relation5point}
to express the full amplitude~\eqref{tree5point} in terms of
two primitives with legs $\u{1}$, $\o{2}$ and $\u{3}$ fixed:
\beal
   {\cal A}_{5,2}^\text{tree} =
      \Big( T_{i_1 \bar \imath_2}^b
            T_{i_3 \bar \jmath}^{a_5} T_{j\,\bar \imath_4}^b
          + T_{i_1 \bar \jmath}^b T_{j\,\bar \imath_2}^{a_5}
            T_{i_3 \bar \imath_4}^b \frac{s_{35}-m_3^2}{s_{25}-m_2^2}
      \Big) & A_{\u{1}\o{2}\u{3} 5 \o{4}} \\
    - \Big( T_{i_1 \bar \jmath}^{a_5} T_{j\,\bar \imath_2}^b
            T_{i_3 \bar \imath_4}^b
          + T_{i_1 \bar \jmath}^b T_{j\,\bar \imath_2}^{a_5}
            T_{i_3 \bar \imath_4}^b \frac{s_{15}-m_1^2}{s_{25}-m_2^2}
      \Big) & A_{\u{1}\o{2}\u{3}\o{4} 5} \,.
\label{tree5final}
\eeal

%%%%%%%%%%%%%%%%%%%%%%%%%%%%%%%%%%%%%%%%%%%%%%%%%%%%
\subsection{Kinematic algebra for $n=6$, $k=3$}
\label{sec:6point}
%%%%%%%%%%%%%%%%%%%%%%%%%%%%%%%%%%%%%%%%%%%%%%%%%%%%

Let us now consider the kinematic structure of the six-quark example
of \sec{sec:multiquark}.
The general massive forms for three of the numerators are found in \eqn{numerators6point},
and in the limit of massless quarks these simplify to
\begin{subequations} \begin{align}
 & n_1(\u{1}^-\!,\o{2}^+\!,\u{3}^-\!,\o{4}^+\!,\u{5}^-\!,\o{6}^+) =
      -i \braket{13} \bra{5}1\!+\!3|4] [62] \,, \\
 & n_2(\u{1}^-\!,\o{2}^+\!,\u{3}^-\!,\o{4}^+\!,\u{5}^-\!,\o{6}^+) =
      -i \braket{15} \bra{3}1\!+\!5|6] [42] \,, \\
 & \begin{aligned} \!
   n_7(\u{1}^-\!,\o{2}^+\!,\u{3}^-\!,\o{4}^+\!,\u{5}^-\!,\o{6}^+) =
      -i \Big( \braket{13} [42] \bra{5}1\!+\!2|6] & 
             + \braket{35} [64] \bra{1}3\!+\!4|2] \\ &
             + \braket{15} [62] \bra{3}5\!+\!6|4] \Big) \,.
   \end{aligned}
\end{align} \label{masslessnumerators6point}% 
\end{subequations}
The remaining four numerators are obtained by permuting the particle labels of these expressions. 
As is evident from this form, it so happens that all the Feynman diagrams are
separately gauge-invariant. Indeed, not only is the quartic vertex absent,
but furthermore there are no external gluons to introduce gauge dependence (\eg via a reference axial vector), as it happens in the five-point amplitude.
Therefore the numerators are unique, and hence the color-kinematics duality cannot be imposed unless it holds from the start:
\beal
   c_1 - c_2 = -c_7 \,, \qquad \qquad \qquad
   n_1 - n_2 = -n_7 \,, \\
   c_3 - c_4 = -c_7 \,, \qquad \qquad \qquad
   n_3 - n_4 = -n_7 \,, \\
   c_5 - c_6 = -c_7 \,, \qquad \qquad \qquad
   n_5 - n_6 = -n_7 \,.
\label{duality6point}
\eeal
Fortunately, these numerator identities do hold,
as can be checked for the massive numerators~\eqref{numerators6point} through Dirac spinor algebra, or in the massless limit (\ref{masslessnumerators6point}) using spinor-helicity identities.

If we now return to the formal definitions of the primitives~\eqref{primitives6point},
and consider them together with the kinematic algebra of \eqn{duality6point}, we obtain
a system of seven equations for seven numerator variables,
which is straightforward to solve uniquely. The first six numerators are given by
\begin{align}
   n_1 = & D_1
         \Big\{ \Big( \frac{1}{D_3\!+\!D_4}
                    + \frac{1}{D_5\!+\!D_6} - \frac{1}{D_7} \Big) n_7
              - A_{\u{1}\o{2}\u{5}\o{6}\u{3}\o{4}}
              + \frac{D_4}{D_3\!+\!D_4} A_{\u{1}\o{2}\u{3}\u{5}\o{6}\o{4}}
              + \frac{D_6}{D_5\!+\!D_6} A_{\u{1}\o{2}\u{5}\u{3}\o{4}\o{6}}
         \Big\} \,, \nn \\
   n_2 = & \: n_1 + n_7 \,, \qquad \qquad
   n_3 =  - \frac{D_3}{D_3\!+\!D_4}
            (D_4 A_{\u{1}\o{2}\u{3}\u{5}\o{6}\o{4}} + n_7) \,, 
\label{solvednumerators6point} \\
   n_4 = & \: n_3 + n_7 \,, \qquad \qquad
   n_5 =  - \frac{D_5}{D_5\!+\!D_6}
            (D_6 A_{\u{1}\o{2}\u{5}\u{3}\o{4}\o{6}} + n_7) \,, \qquad \quad
   n_6 = n_5 + n_7 \,, \nn
\end{align}
and the seventh numerator is
\small
\beal
   n_7 = \Big\{ (D_3\!+\!D_4)(D_5\!+\!D_6)(D_2 A_{\u{1}\o{2}\u{3}\o{4}\u{5}\o{6}} - D_1 A_{\u{1}\o{2}\u{5}\o{6}\u{3}\o{4}})
                -(D_5\!+\!D_6)(D_2 D_3 - D_1 D_4) A_{\u{1}\o{2}\u{3}\u{5}\o{6}\o{4}} & \\
                -(D_3\!+\!D_4)(D_2 D_5 - D_1 D_6) A_{\u{1}\o{2}\u{5}\u{3}\o{4}\o{6}} & \Big\} \\
       / \Big\{  (D_1\!+\!D_2)(D_3\!+\!D_4)(D_5\!+\!D_6)/D_7
                -(D_3\!+\!D_4)(D_5\!+\!D_6) & \\
                -(D_1\!+\!D_2)(D_3\!+\!D_4\!+\!D_5\!+\!D_6) & \Big\} \,.
\eeal
\normalsize
Indeed, we have now expressed all the $n_i$ in terms of gauge-invariant quantities,
formally confirming the direct observation that each Feynman diagram is gauge-invariant. The above expressions look non-local, but by uniqueness of the solution, the numerators have to be local, as is manifest in \eqns{numerators6point}{masslessnumerators6point}.

An interesting fact we learn from this solution is that
there are no extra relations generated\footnote{This does not imply that further amplitude relations do not exist. However, one should expect that any extra relations require more that just the color properties~\eqref{signflip}, \eqref{coloralgebra} and their kinematic analogue~\eqref{duality}.} among the primitive amplitudes in the color-algebra basis~\eqref{primitives6point}.
More generally, this curious absence of BCJ relations happens when there are no external gluons, $n=2k$, as we have explicitly checked up to eight points. At eight points this fact is somewhat counterintuitive since the Feynman graph expansion involves the quartic gluon vertex that naively should lead to numerator ambiguities and associated BCJ relations.

%%%%%%%%%%%%%%%%%%%%%%%%%%%%%%%%%%%%%%%%%%%%%%%%%%%%
\subsection{BCJ relations for QCD}
\label{sec:bcj}
%%%%%%%%%%%%%%%%%%%%%%%%%%%%%%%%%%%%%%%%%%%%%%%%%%%%

In the previous sections we exemplified that scattering amplitudes in QCD
obey the color-kinematics duality. For these examples,
with many quarks and no quartic gluon vertices, it turned out
that the color-kinematics duality followed directly from the Feynman rules.
However, in general this is not the case,
and the duality has to be imposed by hand
using nontrivial rearrangements of terms between the kinematic numerators.
This is what happens for higher-point amplitudes
with only gluons~\cite{Bern:2008qj} or with one quark-antiquark pair
($n \ge 5, k \le 1$). For such amplutudes the numerators satisfying the color-kinematics duality were shown to exist to all multiplicities~\cite{Kiermaier,BjerrumBohr:2010hn,Mafra:2011kj}, thus effectively proving the duality at tree level for $k \le 1$.

By explicit calculations we have checked that the duality works for any quark-gluon configurations up to eight particles. While we do not provide a proof for $n>8, k>1$, we will show that the duality imposes the BCJ relations in QCD that are constitute a well-defined subset of the pure-gluon BCJ relations. 

In the pure-gluon case, the number of independent BCJ relations are $(n-2)!-(n-3)!$ and they are given in ref.~\cite{Bern:2008qj}. A subset of those are linear in momentum invariants,
\be
   \sum_{i=2}^{n-1} \Big( \sum_{j=2}^{i} s_{jn} \Big)
      A(1, 2, \dots i,n,i+1, \dots ,n-1) = 0 \,,
\label{fBCJagain}
\ee
and in \rcite{Feng:2010my} it was shown that relabelings of these simple equations could be used to derive the more complicated relations involving higher powers of momentum invariants. 

We found that the corresponding quark-gluon BCJ relations, for $k$ massive quark lines, are given by the general formula
\be
   \sum_{i=2}^{n-1} \Big( \sum_{j=2}^{i} s_{jn} - m _j^2 \Big)
      A(1, 2, \dots i,n,i+1, \dots ,n-1) = 0 \,.
\label{fBCJmassive}
\ee
where particle $n$ is strictly a gluon, while the remaining $(n-1)$ particles can be of any type: quark/antiquark/gluon. In the next section we will generalize this formula to include the relations with higher powers of momentum invariants, as well as derive the number of linearly independent BCJ relations, counted for $n \leq 8$ in \tab{tab:BCJrelations}.

In \sec{sec:5point} we derived a $n=5$, $k=2$ amplitude relation~\eqref{relation5point} with the permuted leg $n$ being a gluon. It has precisely the form~\eqref{fBCJmassive}. The four-point relation~\eqref{relation4point} from \sec{sec:4point} does not have this precise form right away, it can be easily rewritten that way, either as a sum over different insertions of the gluon leg $4$,
\be
   (s_{24}-m^2) A_{\u{1} \o{2} 4 3}
 + (s_{24} + s_{34} - m^2) A_{\u{1}\o{2} 3 4} = 0 \,,
\ee
or, equivalently, as a sum over insertions of the gluon leg $3$,
\be
   (s_{23}-m^2) A_{\u{1}\o{2} 3 4}
 + (s_{23} + s_{34} - m^2) A_{\u{1}\o{2} 4 3} = 0 \,.
\ee

For the pure-quark six-point amplitude in \sec{sec:6point}, we note that \eqn{fBCJmassive} is consistent with the fact it had no BCJ amplitude relations despite obeying the color-kinematics duality. Indeed, by definition pure-quark amplitudes, $n=2k$, have no external gluons, and thus \eqn{fBCJmassive} gives no relations for them.

%%%%%%%%%% TABLE %%%%%%%%%%
\begin{table*}
\centering
\begin{tabular}{|c||c|c|c|c|c|c|}
\hline
$k \setminus n$ & 3 & 4 & 5 & 6 & 7 & 8 \\
\noalign{\hrule height 0.7pt}
0 & 0 & 1 & 4 & 18 & 96 & 600 \\
\hline
1 & 0 & 1 & 4 & 18 & 96 & 600 \\
\hline
2 & - & 0 & 1 &  6 & 36 & 240 \\
\hline
3 & - & - & - &  0 &  4 &  40 \\
\hline
4 & - & - & - &  - &  - &   0 \\
\hline
\end{tabular}
\\ \vspace{8pt}
$ \delta(n,k)=
  \left\{ \begin{array}{ll}
          (n-2)! - (n-3)! ~&~ \text{for } k = 0,1 \\
          \big( (n-2)! - (n-3)!(2k-2) \big)/k! ~&~ \text{for }2 < 2k \le n \\
          \end{array} \right. $
\\ \vspace{-7pt}
\caption{\small Number of independent BCJ relations, $\delta(n,k)$,
         for $n$-point amplitudes with $k$ distinguishable quark pairs
         and $(n-2k)$ gluons.}
\label{tab:BCJrelations}
\end{table*}
%%%%%%%%%%%%%%%%%%%%%%%%%%%

We derived the quark-gluon BCJ relations~\eqref{fBCJmassive} as follows.
We start with the primitive amplitudes in the Melia basis~\eqref{KKbasis},
\be
   \Big\{ A(\u{1},\o{2},\sigma) = \!\!\!
          \sum_{\sigma\text{-color-ordered cubic graphs }\Gamma_i} \!\!\!
          \pm\,\frac{n_i}{D_i} ~~~\Big|~~~
          \sigma \in \text{Melia basis} \Big\} \,,
\label{MeliaBasisGraphs}
\ee
written in terms of the formal numerators $n_i$, and the given kinematic denominators $D_i$.
The numerators are further constrained by the kinematic Jacobi/commutation relations:\footnote{The
number of independent Jacobi/commutation relations is\;
$\nu(n,k)-\varkappa(n,k) = \frac{(2n-5)!!}{(2k-1)!!}-\frac{(n-2)!}{k!}$.}
\be
   \Big\{ n_i - n_j = n_k ~~\Big|~~
          (i,j,k) \in \text{Jacobi/commutation triplets} \Big\} \,.
\label{JacobiTriplets}
\ee
Solving the combination of the systems \eqref{MeliaBasisGraphs} and \eqref{JacobiTriplets}, similar to how it was done in sections \ref{sec:4point}--\ref{sec:6point}, one obtains a collection of gauge-dependent solutions for a subset of the $n_i$ and a collection of gauge-invariant solutions for a subset of primitives. The latter solutions are the BCJ relations.

We explicitly solved this system for any quark-gluon configuration
up to eight particles, as well as for the nine-point amplitude with four quark lines and one gluon. 
We also verified that, similarly to the pure-gluon case,
all the BCJ relations in QCD follow from label permutations of the simple family of relations of the type~\eqref{fBCJmassive}.
However, the label permutations of \eqn{fBCJmassive} should be regarded as an equation system, not a solution, to the full set of BCJ relations. In the full solution the primitive amplitudes are reduced to a new basis smaller than that of the color-algebra basis~\eqref{KKbasis}.

%%%%%%%%%%%%%%%%%%%%%%%%%%%%%%%%%%%%%%%%%%%%%%%%%%%%
\subsection{New amplitude basis for QCD}
\label{sec:bcjsolution}
%%%%%%%%%%%%%%%%%%%%%%%%%%%%%%%%%%%%%%%%%%%%%%%%%%%%

Now let us find a complete solution to all gauge-invariant relations
imposed by the system of equations
comprised of \eqns{MeliaBasisGraphs}{JacobiTriplets}.
These relations are the BCJ relations and allow us to reduce the primitive amplutudes to a smaller basis. 

The new BCJ basis for $k\ge2$ is taken to be a subset
of the Melia basis~\eqref{KKbasis} obtained by demanding that
the third particle is strictly a quark:
\be
   \big\{ A(\u{1},\o{2},\u{q},\sigma) ~\big|~
          \{\u{q},\sigma\} \in \text{Dyck}_{k-1}
          \times \{\text{gluon insertions in}~\sigma\}_{n-2k} \big\} \,.
\label{BCJbasis}
\ee
It is always possible to demand that for $k>1$, because any gluon that happens to be in the third location can be moved to other positions using \eqn{fBCJmassive}.

Let us count the size of the basis \eqref{BCJbasis}. As before, there are $(2k-2)!/(k!(k-1)!)$ Dyck words, and after dressing them with quark labels, each allow for $(k-1)!$ different quark pair configurations.
The $(n-2k)$ gluons are then free to be assigned to the slots in-between the particles, except for the space inside the fixed sequence $\u{1},\o{2},\u{q}$.
For each gluon inserted the available number of slots increases, ranging from $(2k-2)$ to $(n-3)$.
Thus the length of the basis is
\be
   \beta(n,k) =
   \underbrace{ \overbrace{\frac{(2k-2)!}{k!(k-1)!}}^\text{empty brackets}
                \times (k-1)!}_\text{dressed quark brackets}
      \times \underbrace{(2k-2)(2k-1)\dots(n-3)}_{\text{insertions of }
                                                  (n-2k)\text{ gluons}}
      = \frac{(n-3)!(2k-2)}{k!} \,.
\label{BCJbasisLength}
\ee

For the amplitudes that only have one quark/antiquark pair, $k=1$, it is not possible to pick the third particle to be a quark.  So instead we pick it to be the gluon $3$, giving the basis $A(\u{1},\o{2},3,\sigma)$. Except for the bars on the labels, this is the same basis as for the pure-gluon case~\cite{Bern:2008qj}, thus the basis for $k=0,1$ is of size $(n-3)!$.
The basis counts for different quark-gluon configurations
are exemplified and summarized in \tab{tab:IndependentBCJ}.

%%%%%%%%%% TABLE %%%%%%%%%%
\begin{table*}
\centering
\begin{tabular}{|c||c|c|c|c|c|c|c|c|}
\hline
$k \setminus n$ & 3 & 4 & 5 & 6 & 7 & 8 \\
\noalign{\hrule height 0.7pt}
0 & 1 & 1 & 2 & 6 & 24 & 120 \\
\hline
1 & 1 & 1 & 2 & 6 & 24 & 120 \\
\hline
2 & - & 1 & 2 & 6 & 24 & 120 \\
\hline
3 & - & - & - & 4 & 16 &  80 \\
\hline
4 & - & - & - & - &  - &  30 \\
\hline
\end{tabular}
\\ \vspace{8pt}
$ \beta(n,k)=
  \left\{ \begin{array}{ll}
          (n-3)! ~&~ \text{for }k = 0,1 \\
          (n-3)!(2k-2)/k! ~&~ \text{for }2 < 2k \le n \\
          \end{array} \right. $
\\ \vspace{-7pt}
\caption{\small Number of independent primitive amplitudes, $\beta(n,k)$,
         in the full $n$-point amplitude with $k$ distinguishable quark pairs
         and $(n-2k)$ gluons, after imposing the BCJ relations.}
\label{tab:IndependentBCJ}
\end{table*}
%%%%%%%%%%%%%%%%%%%%%%%%%%%

The full solution to the BCJ relations are given by first solving the numerators $n_i$ in terms of the primitives in the BCJ basis \eqref{BCJbasis}, and then plugging them into the primitives that are not part of this basis.  Since the color-algebra basis~\eqref{KKbasis} already has
legs $\u{1}$ and $\o{2}$ next to each other,
we only need to give the reduction formula for the primitives with
the first quark leg $\u{q}$ separated from leg $\o{2}$
by a set of gluonic legs $\alpha$
and followed by a mixed set of quark-gluon legs $\beta$:
\be
   A(\u{1},\o{2},\alpha,\u{q},\beta) \,.
\ee
To simplify the subsequent formulas, we choose $\u{q}=\u{3}$ and
take the leg labels in sets $\alpha$ and $\beta$
to be consecutive numbers:
\begin{equation}
   \alpha \equiv \{4,5,\ldots,p-1,p\}\,, \qquad
   \u{q} \equiv \u{3}\,, \qquad
   \beta \equiv \{p+1,p+2,\ldots,n-1,n\} \,. 
   \label{aqb}
\end{equation}
As already mentioned, $\alpha$ consists strictly of gluon legs, and the particles in $\beta$  can be of any type: quark/antiquark/gluon.
The consecutive labeling choice can always be undone in the final expressions
by relabeling of legs $3,4,\ldots,n$.

By extrapolating from the structure
of the solutions evaluated up through eight external particles,
we obtain the following all-multiplicity formula:
\begin{equation}
   A(\u{1},\o{2},\alpha,\u{3},\beta) = \!\!
      \sum_{\sigma \in S(\alpha) \shuffle \beta} \!\!\!
      A(\u{1},\o{2},\u{3},\sigma) \prod_{i=4}^p
      \frac{ {\cal F}(3,\sigma,1|i) }{ s_{2,4,\ldots,i} - m_2^2 } \,,
\label{alln}
\end{equation}
where the sum runs over partially ordered permutations of
the merged sets $\alpha$ and $\beta$.
The shuffle product is used in $S(\alpha) \shuffle \beta$ to obtain all
permutations of $\alpha \cup \beta$
that preserve the order of the elements of $\beta$. Note that such permutations maintain the valid bracket structure of the quark labels. 
The kinematic function ${\cal F}$ associated with the gluon leg $i$ is given by
\beal
{\cal F}(3,&\,\sigma_1,\sigma_2, \ldots, \sigma_{n-3}, 1 | i) 
\equiv {\cal F} (\rho|i) \\
=&
\left\{
\begin{array}{ll}
      \sum_{l=t_i}^{n-1} {\cal G}(i,\rho_l) ~&~ \mbox{if $t_{i-1} < t_i$}\\
    - \sum_{l=1}^{t_i} {\cal G}(i,\rho_l) ~&~ \mbox{if $t_{i-1} > t_i$}
\end{array} \right\}
+
\left\{
\begin{array}{ll}
      s_{2,4,\ldots,i} - m_2^2 ~&~ \mbox{if $t_{i-1} < t_i< t_{i+1}$}\\
    - s_{2,4,\ldots,i} + m_2^2 ~&~ \mbox{if $t_{i-1} > t_i > t_{i+1}$}\\
      0 ~&~ \mbox{else}
\end{array} \right\} \,, \!\!
\label{Ffunction}
\eeal
and where $t_k$ is the position of leg $k$ in the set $\rho$,
except for $t_3$ and $t_{p+1}$, which are always defined to be\footnote{An
alternative choice is  $t_3\equiv \infty, t_{p+1} \equiv 0$,
which is equivalent to \eqn{BoundaryConditions} by momentum conservation.}
\begin{equation}
   t_3 \equiv t_5 \,, \qquad \quad
   t_{p+1} \equiv 0 \,.
\label{BoundaryConditions}
\end{equation}
(Note that for $p=4$ this implies $t_3=t_{p+1}=0$.)
The function ${\cal G}$ is given by
\begin{equation}
{\cal G}(i,j)=\left\{
\begin{array}{ll}
      s_{ij} - m_j^2 ~&~ \mbox{if $i < j$ or $j=1,3$}\\
      0 ~&~ \mbox{else}
\end{array} \right\} \,.
\label{Gdef}
\end{equation}

For the case that $\alpha$ consists of a single gluon,
the formula~\eqref{alln} becomes equivalent to \eqn{fBCJmassive}.
If the amplitude has a single massive quark line, $k=1$, the relations~\eqref{alln} still hold after converting leg $\u{q}$ to a gluon: $\u{3}\rightarrow 3$. Similarly, in the pure-gluon case, dropping all bars reduces~\eqn{alln} to the standard BCJ relations~\cite{Bern:2008qj}. We have explicitly checked that all QCD amplitudes through eight points satisfy the BCJ relations~\eqref{alln}.

A known feature of the BCJ relations is that they are very general. We expect~\eqn{alln} to be valid not only for ordinary QCD, but for more generic gauge theories describing the interactions of adjoint vectors and massive matter transforming in any representation of the gauge group. In particular, supersymmetric versions of QCD, as well as $D$-dimensional extensions of QCD, are examples of theories that should obey~\eqn{alln}. At tree level there is little difference between super-QCD amplitudes and those of QCD. The reason is that the gluons and quarks directly translate to supersymmetric adjoint and fundamental multiplets, and the structure of the supersymmetric tree-level interactions are dictated by those of the gluon and quark states (which are top components of respective supermultiplets).

The pure-gluon BCJ relations are known to be valid in any space dimension~\cite{Bern:2008qj,BjerrumBohr:2009rd,Stieberger:2009hq,Feng:2010my}, and we find the same is true for the quark-gluon amplitudes considered in this paper. For example, we used only $D$-dimensional properties to show that the numerators generated from the Feynman rules in sections~\ref{sec:4point}--\ref{sec:6point} obey the color-kinematics duality. The observation that it holds for $D$-dimensional quark-gluon amplitudes explains why the generalization to massive (Dirac) quarks is straightforward. As is well known, the mass of the quarks can be reinterpreted as a component of a higher-dimensional massless momenta, living in some an internal space (\ie similar to the Kaluza-Klein construction).  Indeed, the factors $(s_{2,4,\dots,i} - m_2^2)$ and $(s_{ij} - m_j^2)$ that appear in eqns.~(\ref{alln}--\ref{Gdef}) can be thought of as being higher-dimensional versions of $s_{2,4,\dots,i}$ and $s_{ij}$, given that the quarks have momenta $(k_i^\mu,m_i)$, antiquarks momenta $(k_i^\mu,-m_i)$, and gluons have momenta $(k_i^\mu,0)$.

%%%%%%%%%%%%%%%%%%%%%%%%%%%%%%%%%%%%%%%%%%%%%%%%%%%%
\subsection{New mixed decomposition}
\label{sec:ckfactors}
%%%%%%%%%%%%%%%%%%%%%%%%%%%%%%%%%%%%%%%%%%%%%%%%%%%%

Finally,
we can combine our color decomposition~\eqref{GeneralQuarkGluonDecomposition}
with the solution to the BCJ relations to decompose the full amplitude
in the BCJ basis of primitives~\eqref{BCJbasis}.

For simplicity, let us start with the case of $k=0,1$.
We can rewrite both \Eqns{QuarkLineDecomposition}{DDM} as
\be
    {\cal A}^{\text{tree}}_{n,k} = \!\!
      \sum_{\sigma \in S_{n-2}(\{3,\dots,n\})} \!\!\!
      C(1,2,\sigma)\,A(1,2,\sigma) \,,
\label{DDMimplicit}
\ee
where
$ C(1,2,\sigma) =
  \big( T^{a_{\sigma(3)}} \dots T^{a_{\sigma(n)}} \big)_{\bar \jmath_2 i_1} $
in the fundamental or adjoint representation.
The sum over permutations of arguments $3,\dots,(n-3)$ can be divided into those terms with
with particles $1,2,3$ appear together, and those terms where $2$ and $3$ are separated by $\alpha$,
\be \!\!\!\!
    {\cal A}^{\text{tree}}_{n,k} = \!\!\!\!
      \sum_{\sigma \in S_{n-3}(\{4,\dots,n\})} \!\!\!\!\!
         C(1,2,3,\sigma)\,A(1,2,3,\sigma) \, + \!\!\!\!\!\!\!
      \sum_{\substack{~~\alpha \cup \beta = \{4,\dots,n\} \\
                        \alpha \cap \beta =\,\emptyset~~~~\; }}\!\!\!\!
         C(1,2,\alpha,3,\beta)\,A(1,2,\alpha,3,\beta) \,. \!\!
         \label{decompBCJ}
\ee
The BCJ relations~\cite{Bern:2008qj} allow us to convert all the primitives
in the second sum to the BCJ basis, with the three first legs fixed,
\be
   A(1,2,\alpha,3,\beta) = \!\!
      \sum_{\sigma \in S(\alpha) \shuffle \beta} \!\!\!
      A(1,2,3,\sigma)
      \prod_{i =1}^{|\alpha|}
      \frac{ {\cal F}(3,\sigma,1| i) }
           { s_{2,\alpha_1,\ldots,\alpha_i} - m_2^2 } \,,
\label{BCJsolution}
\ee
where for non-canonical orderings the function ${\cal F}$ is defined via label permutations of \eqn{Ffunction}.
By construction, since the permutation $\sigma$ is an element in $S(\alpha) \shuffle \beta$, it follows that $\beta$ is a subset of $\sigma$. Hence, after plugging \eqn{BCJsolution} into \eqn{decompBCJ}, one can rearrange the order of summations by defining $\beta \subset \sigma$ and $\alpha \in S(\sigma \setminus\beta)$. This gives the BCJ decomposition
\begin{align}
\label{BCJdecomposition}
    {\cal A}^{\text{tree}}_{n,k\le1} = \!\!
    \sum_{\alpha \in S_{n-3}(\{4,\dots,n\})} \!\!\!\!
      A(1,2,3,\sigma) & \\ \times
    \bigg\{
      C(1,2,3,\sigma) & \,
    + \sum_{\beta \subset \sigma} \;
      \sum_{\alpha \in S(\sigma\setminus\beta)} \!\!
      C(1,2,\alpha,3,\beta)
      \prod_{i =1}^{|\alpha|}
      \frac{ {\cal F}(3,\sigma,1| i) }
           { s_{2,\alpha_1,\ldots,\alpha_i} - m_2^2 }
    \bigg\} \,. \nn
\end{align}
This expresses the full amplitude in terms of the basis of primitives
independent under both KK and BCJ relations.
Note that their gauge-invariant coefficients now contain
not only color but also kinematic factors.

It is now straightforward to generalize
the decomposition~\eqref{BCJdecomposition} to the general quark-gluon case.
For that, the external sum should run over the BCJ basis~\eqref{BCJbasis},
in which the quarks $\u{q}$ directly succeed $\u{1},\o{2}$,
while the internal sum over $\beta$ should include
all quark and antiquark labels, and thus the complement set $\sigma\setminus\beta$ should be purely gluonic:
\begin{align}
\label{BCJdecomposition2}
    {\cal A}^{\text{tree}}_{n,k\ge2} = \!\!
    \sum_{(\u{q},\sigma)\:\!\in\text{\,BCJ basis}} \!\!\!
      A(\u{1},\o{2},\u{q},\sigma) & \\ \times
    \bigg\{
      C(\u{1},\o{2},\u{q},\sigma) & \,
    + \sum_{\substack{ \beta \subset \sigma \\
                       \sigma\setminus\beta~\text{gluonic}}} \;
      \sum_{\alpha \in S(\sigma\setminus\beta)} \!\!
      C(\u{1},\o{2},\alpha,\u{q},\beta)
      \prod_{i =1}^{|\alpha|}
      \frac{ {\cal F}(q,\sigma,1| i) }
           { s_{2,\alpha_1,\ldots,\alpha_i} - m_2^2 }
    \bigg\} \,, \nn
\end{align}
where the color structures are defined as in \eqn{KKcolor}.
This decomposition should hold for any $k=2,\dots, \lfloor n/2 \rfloor$. In fact, it also holds in the $k=0,1$ case. Indeed, once the bars are dropped and $\u{q} \rightarrow 3$, it becomes indistinguishable from \eqn{BCJdecomposition}.
In the pure-quark case, $n=2k$, the internal sum over the gluon set~$\alpha$ vanishes,
and the BCJ basis collapses to the Melia basis,
making \eqn{BCJdecomposition2} equivalent to the color decomposition~\eqref{NewColorDecomposition}.

We have explicitly checked the decompositions~(\ref{BCJdecomposition}) and (\ref{BCJdecomposition2}) through eight points,
as well as for the nine-point case with four quark-antiquark pairs.

\section{Summary and discussion}
\label{sec:conclusion}

In this paper we explored and organized the color and kinematic content
of general tree amplitudes in QCD with flavored massive quarks and massless gluons. We decomposed the amplitudes into reduced sets of color-ordered primitive amplitudes and associated color coefficients, and found that the amplitudes obey the color-kinematics duality.

A familiar observation is that the true color space of a gauge-theory amplitude
is smaller than the space spanned by the classic SU($N_c$) color basis expansion. Exploiting this fact, we obtained a new color decomposition for $n$-point QCD amplitudes with $k$ quark-antiquark pairs, which involves only the $(n-2)!/k!$ planar color-stripped primitives that belong to the Melia basis~\cite{Melia:2013bta,Melia:2013epa}.
This decomposition can be regarded as a natural $(n,k)$-generalization
of the known Del Duca-Dixon-Maltoni decomposition for gluons~\cite{DelDuca:1999ha,DelDuca:1999rs}, which uses only the $(n-2)!$ color-ordered primitives comprising the Kleiss-Kuijf basis~\cite{Kleiss:1988ne}. 

The DDM decomposition is advantageous because of its analytic compactness.
For example, it has been successfully used for efficient computations of the color-subleading parts of QCD loop amplitudes~\cite{Badger:2012pg,Badger:2015lda}. Similarly, one can hope that the new decomposition can be used to improve on the efficiency of calculations relevant for LHC phenomenology. However, for calculations of color-averaged cross-sections, properties other than those sought after in formal amplitude calculations may become more important. For instance, it can be more convenient to have color factors that are orthogonal to each other (\eg see the recent \rcite{Du:2015apa}). The color factors used in the decomposition presented here consist of those appearing in the Feynman diagrams of the amplitude, thus they are completely general and can encode any gauge group and any group representation.

The new color decomposition has an interesting hierarchy of nested structures of quark lines. For the least nested configurations the color factors are planar and similar to the multi-peripheral diagrams, whereas the color coefficients of the more nested primitives have an intricate nonplanar (with respect to particle ordering) structure.
This structure is controlled by multi-tensor representations of the gauge group, with generators $\Xi^a_l$, suggesting that a deeper understanding of the Lie-algebraic structure of the decomposition could be advantageous. Using the higher-representation generators $\Xi^a_l$, we obtain surprisingly compact expressions for the color coefficients of the primitive amplitudes, which can be rather formidable expressions when expressed in terms of ordinary color factors. The color diagrams that we use for the color coefficients generalize the multi-peripheral diagrams used in the DDM decomposition. 

Nested color structures of quark lines and gluons also appear in other gauge-invariant observables, \eg in processes where massive quarks and hard gluons can be approximated by Wilson lines. These are important for understanding the soft and collinear singularity structure of gauge-theory amplitudes~\cite{Dixon:2009ur}. Non-abelian exponentiation theorems allow the Wilson line calculations to be reduced to simpler web diagrams.  Similar to the amplitudes considered in this paper, the color factors arising from webs are not independent, and the mixing of color and kinematic degrees of freedom, which
plays an essential role for the infrared exponentiation~\cite{Gardi:2010rn,Gardi:2011yz,Gardi:2013ita},
is governed by nontrivial combinatorics~\cite{Dukes:2013gea}.
It would be interesting to better understand the common aspects of the new color decomposition and well-known properties of Wilson lines and webs.

In this paper we provided substantial evidence that the color-kinematics duality~\cite{Bern:2008qj,Bern:2010ue} is present in QCD. In particular, we considered the duality for the first time in the context of arbitrary numbers of distinctly flavored massive fundamental Dirac fermions (quarks), and show that they mesh well with it. By explicit calculations up to eight particles (quarks and gluons), we showed that one can find kinematic numerators that obey the same commutation/Jacobi identities as the corresponding color factors. This suggests that a kinematic algebra, analogous to the color algebra, controls the QCD amplitudes. We do not explicitly consider loop amplitudes in this paper, but given the existence of the duality at tree level we conjecture that it should also be present in loop amplitudes of QCD. Indeed, through the unitarity method~\citeUnitarityMethod~we know that any unitarity cut, which breaks up a loop amplitude into products of tree amplitudes, will inherit the color-kinematics duality of the corresponding trees. Every such unitarity cut should then have a diagrammatic expansion where the numerators obey the color-kinematics duality, which is highly suggestive of a kinematic algebra that exists at loop level. Certain simple massless QCD amplitudes have already been shown to obey the color-kinematics duality at one and two loops~\cite{Boels:2013bi,Bjerrum-Bohr:2013iza,Bern:2013yya,Nohle:2013bfa,Johansson:2014zca}, adding to the credibility of the conjecture. 

Using the color-kinematics duality, we derived a complete set of gauge-invariant BCJ amplitude relations for QCD at tree level. These amplitude relations are massive $(n,k)$-generalizations of the purely gluonic BCJ relations~\cite{Bern:2008qj}. While the analytic form of the new relations is similar to the pure-gluon relations, their implications differ substantially. In particular, the new relations clarify the different roles that gluons (vectors) and quarks (matter) play in the context of the color-kinematics duality. The existence of the BCJ relations is directly tied to the presence of external gluons. Pure-quark amplitudes have no BCJ relations, while pure-gluon amplitudes have the maximum number of such relations. The most elementary type of BCJ relations corresponds to the action of moving a single gluon around and inserting it at possible locations along the color-ordered amplitude. This is similar to the action of a Ward identity, or, indeed, the action of a Lie algebra. That BCJ relations arise only from the presence of external gluons supports the observation that these relations are intimately tied to the gauge freedom of the theory~\cite{Bern:2008qj,Bern:2010ue,Bern:2010yg}.

Using the new BCJ amplitude relations, we found a new basis of $(n-3)!(2k-2)/k!$ primitives for $k\ge2$, which should be contrasted to the well-known BCJ basis of size $(n-3)!$ for pure-gluon or single-quark-line amplitudes, $k=0,1$. The new basis allows us to construct an amplitude decomposition that uses only these primitives. In doing so, we introduce a mixture of kinematic dependence and dependence on color factors into the coefficients of the primitives. It is interesting to note that for $k\ge2$ this decomposition sidesteps $(n-2k)(n-3)!/k!$ of the Melia basis primitives, and for $k=0,1$ it sidesteps $(n-3)(n-3)!$ of the Kleiss-Kuijf basis primitives. For the reader's convenience,
a Mathematica implementation of the amplitude decompositions is provided in the ancillary file~\cite{Aux}.

Finally, we note that due to the duality of color and kinematics one should be able to swap color structures for kinematic structures in several formulas of this paper, since they obey the same general Lie-algebraic relations, from which most results were derived. For example, if one takes the new color decomposition introduced in \sec{sec:colorfactors} and replaces the color coefficients $C(1,2,\sigma)$ with the corresponding kinematic coefficients $K(1,2,\sigma)$, it should give a gravitational scattering amplitude
\be
   {\cal M}_{n,k}^{\text{tree}} = \sum_{\sigma\in \text{Melia basis}}
      K(1,2,\sigma) A(1,2,\sigma) \,,
\ee
where the gravitational coupling has been suppressed.
Analogous to the color coefficient, $K(1,2,\sigma)$ is a local kinematic function that consists of a sum over kinematic numerators. Like the numerators, this function is generically gauge-dependent. By the color-kinematics duality, $K(1,2,\sigma)$ should have a formula that mimics the formula for $C(1,2,\sigma)$. This suggest that the higher-representation generator $\Xi_l^a$ should have a kinematic analogue, which would represent the action of a gluon on tensor products of fermion lines. Improved understanding of this kinematic object would most likely be a concrete step towards unraveling the kinematic Lie algebra of QCD.

%%%%%%%%%%%%%%%%%%%%%%%%%%%%%%%%%%%%%%%%%%%%%%%%%%%%
\acknowledgments

We would like to thank Simon Badger, Einan Gardi and Donal O'Connell for helpful discussions. We also thank Marco Chiodaroli, Murat G\"{u}naydin, and Radu Roiban for conversations and collaborations on topics related to this work.
We are grateful to John Joseph Carrasco, Lance Dixon, David Kosower
and Gregor K\"alin for discussions and helpful comments on the manuscript. HJ's work is supported in part by the Swedish Research Council under grant 621--2014--5722, the Knut and Alice Wallenberg Foundation under grant KAW~2013.0235 (Wallenberg Academy Fellow), and the CERN-COFUND Fellowship program (co-funded via Marie Curie Actions grant PCOFUND--GA--2010--267194 under the European Union's Seventh Framework Programme). AO's work is supported by the EU via Marie Curie Actions grant FP7-PEOPLE-2013-CIG.

%%%%%%%%%%%%%%%%%%%%%%%%%%%%%%%%%%%%%%%%%%%%%%%%%%%%
\appendix

%%%%%%%%%%%%%%%%%%%%%%%%%%%%%%%%%%%%%%%%%%%%%%%%%%%%
\section{Color-ordered Feynman rules}
%%%%%%%%%%%%%%%%%%%%%%%%%%%%%%%%%%%%%%%%%%%%%%%%%%%%
\label{app:FeynmanRules}

For completeness, in this appendix
we give the color-stripped Feynman rules~\cite{Dixon:1996wi},
which are consistent with the color vertices in \fig{fig:colorvertices}.
All momenta are considered outgoing.
\begin{subequations} \begin{align}
\parbox{45pt}{
\begin{fmffile}{gvertex} \fmfframe(0,10)(0,10){
\fmfset{curly_len}{5pt}
\begin{fmfgraph*}(30,30)
      \fmflabel{$k,\lambda\!$}{g1}
      \fmflabel{$p,\mu$}{g2}
      \fmflabel{$q,\nu$}{g3}
      \fmftop{g2}
      \fmfbottom{g1,g3}
      \fmf{curly}{v,g1}
      \fmf{curly}{v,g2}
      \fmf{curly}{v,g3}
      \end{fmfgraph*} }
      \end{fmffile} }
 & =\; \frac{i}{\sqrt{2}} \big[ g^{\lambda \mu} (k-p)^\nu
                              + g^{\mu \nu} (p-q)^\lambda
                              + g^{\nu \lambda} (q-k)^\mu \big] \,, \\
\parbox{45pt}{
\begin{fmffile}{g4vertex} \fmfframe(0,10)(0,10){
\fmfset{curly_len}{5pt}
\begin{fmfgraph*}(30,33)
      \fmftop{g2,g3}
      \fmfbottom{g1,g4}
      \fmf{curly}{v,g1}
      \fmf{curly}{v,g2}
      \fmf{curly}{v,g3}
      \fmf{curly}{v,g4}
      \fmfv{label=$\lambda$,label.angle=-135,label.dist=5pt}{g1}
      \fmfv{label=$\mu$,label.angle=135,label.dist=5pt}{g2}
      \fmfv{label=$\nu$,label.angle=45,label.dist=5pt}{g3}
      \fmfv{label=$\rho$,label.angle=-45,label.dist=5pt}{g4}
      \end{fmfgraph*} }
      \end{fmffile} }
 & =\; i\:\!g^{\lambda \nu} g^{\mu \rho}
     - \frac{i}{2} \big( g^{\lambda \mu} g^{\nu \rho}
                       + g^{\lambda \rho} g^{\mu \nu} \big) \,, \\
\parbox{45pt}{
\begin{fmffile}{qvertex} \fmfframe(0,10)(0,10){
\fmfsettings
\begin{fmfgraph*}(30,30)
      \fmftop{g1}
      \fmfbottom{q1,q2}
      \fmf{curly,tension=0.91}{v,g1}
      \fmf{plain_arrow}{q2,v,q1}
      \fmfv{label=$\mu$,label.angle=90,label.dist=5pt}{g1}
      \end{fmfgraph*} }
      \end{fmffile} }
 & =\; \frac{i}{\sqrt{2}} \gamma^\mu \,, \qquad \qquad \quad
\parbox{45pt}{
\begin{fmffile}{qbarvertex} \fmfframe(0,8)(0,12){
\fmfsettings
\begin{fmfgraph*}(30,30)
      \fmfbottom{g1}
      \fmftop{q1,q2}
      \fmf{curly,tension=0.91}{v,g1}
      \fmf{plain_arrow}{q2,v,q1}
      \fmfv{label=$\mu$,label.angle=-90,label.dist=5pt}{g1}
      \end{fmfgraph*} }
      \end{fmffile} }
   = \,-\frac{i}{\sqrt{2}} \gamma^\mu \,, \\
\parbox{45pt}{
\begin{fmffile}{gpropagator} \fmfframe(0,10)(0,10){
\fmfsettings
\begin{fmfgraph*}(30,15)
      \fmfleft{g1}
      \fmfright{g2}
      \fmf{curly,label=$p$,label.dist=9pt}{g2,g1}
      \fmfv{label=$\mu$,label.angle=180,label.dist=5pt}{g1}
      \fmfv{label=$\nu$,label.angle=0,label.dist=5pt}{g2}
      \end{fmfgraph*} }
      \end{fmffile} }
 & = \,-\frac{i g_{\mu \nu}}{p^2} \,, \qquad \qquad \quad \;\!\!
\parbox{45pt}{
\begin{fmffile}{qpropagator} \fmfframe(0,10)(0,10){
\fmfsettings
\begin{fmfgraph*}(30,15)
      \fmfleft{q1}
      \fmfright{q2}
      \fmf{plain_arrow,label=$p$,label.dist=6pt}{q2,q1}
      \end{fmfgraph*} }
      \end{fmffile} }
   = \,\frac{i ({\not}p + m)}{p^2 - m^2} \,.
\end{align} \label{kinematicvertices}%
\end{subequations}

%%%%%%%%%%%%%%%%%%%%%%%%%%%%%%%%%%%%%%%%%%%%%%%%%%%%
\bibliographystyle{JHEP}
\bibliography{references}

\end{document}